\documentclass[superscriptaddress, groupedaddress, twocolumn, amsmath,amssymb, aps, prl]{revtex4}
\usepackage{graphicx}
\usepackage{bm}
\begin{document}

\title{Rotating atomic quantum gases with light-induced azimuthal gauge potentials and the observation of the Hess-Fairbank effect}

\author{P.~-K. Chen}
\author{L.~-R. Liu}
\author{M.~-J. Tsai}
\author{N.~-C. Chiu}
\affiliation{Institute of Atomic and Molecular Sciences, Academia Sinica, Taipei, Taiwan 10617}
\author{Y. Kawaguchi}
\affiliation{Department of Applied Physics, Nagoya University, Nagoya, 464-8603, Japan}
\author{S.~-K. Yip}
\affiliation{Institute of Atomic and Molecular Sciences, Academia Sinica, Taipei, Taiwan 10617}
\affiliation{Institute of Physics,Academia Sinica, Taipei, Taiwan 11529}
\author{M.~-S. Chang}
\affiliation{Institute of Atomic and Molecular Sciences, Academia Sinica, Taipei, Taiwan 10617}
\author{Y.~-J. Lin}
\email{linyj@gate.sinica.edu.tw}
\affiliation{Institute of Atomic and Molecular Sciences, Academia Sinica, Taipei, Taiwan 10617}
\date{\today}

\begin{abstract}
We demonstrate synthetic azimuthal gauge potentials for
Bose-Einstein condensates from engineering atom-light couplings. The
gauge potential is created by adiabatically loading the condensate
into the lowest energy Raman-dressed state, achieving a coreless
vortex state. The azimuthal gauge potentials act as effective
rotations and are tunable by the Raman coupling and detuning. We
characterize the spin textures of the dressed states, in agreements
with the theory. The lowest energy dressed state is stable with a
4.5-s half-atom-number-fraction lifetime. In addition, we exploit
the azimuthal gauge potential to demonstrate the Hess-Fairbank
effect, the analogue of Meissner effect in superconductors. The
atoms in the absolute ground state has a zero quasi-angular momentum
and transits into a polar-core vortex when the synthetic magnetic
flux is tuned to exceed a critical value. Our demonstration serves
as a paradigm to create topological excitations by tailoring
atom-light interactions where both types of SO(3) vortices in the
$|\langle \vec{F}\rangle|=1$ manifold, coreless vortices and
polar-core vortices, are created in our experiment. The gauge field
in the stationary Hamiltonian opens a path to investigating rotation
properties of atomic superfluids under thermal equilibrium.
\end{abstract}

\maketitle

Synthetic gauge fields for ultracold neutral
atoms~\cite{lin09,aidelsburger2013,miyake2013,Struck2012,Parker2013,jotzu2014,Dalibard11,Goldman2013,Zhai2015}
marked one of the milestones toward creating topological quantum
matters and exploring novel quantum phenomenon. Among various
implementations, the optical Raman coupling scheme is to couple
different internal spin states while transferring photon momentum.
This leads to the spin-linear-momentum
coupling~\cite{Goldman2013,Zhai2015,Lin11,Wu2016,Huang2016}, which
is a type of ``general spin-orbit-coupling" (SOC), referring to
coupling between the atomic spin and the center-of-mass motion of
the atoms. Another class of SOC where the atomic spin is coupled to
the orbital-angular-momentum (OAM) of the
atoms~\cite{Juzeliunas2005}, has been demonstrated using one
Laguerre-Gaussian (LG) Raman beam carrying OAM of
light~\cite{Chen2018}; this is the spin-orbital-angular-momentum
coupling (SOAMC)~\cite{Qu2015,DeMarco2015,Hu2015,Chen2016}.

In SOAMC systems, the atoms dressed by LG Raman beams experience
azimuthal gauge potentials $\vec{A}=A(r){\mathbf e}_{\phi}$, where
the azimuthal dispersion is $(\ell-\ell_{\rm min})^2/2mr^2$ at
radial position $r$ and $\ell$ is the angular momentum. The shifted
minimum is $\ell_{\rm min}=r A(r)$, where $A(r)$ is tunable by the
Raman coupling strength $\Omega(r)$ and Raman detuning $\delta$ in
our experiment. This azimuthal gauge potential is thus equivalent to
an effective rotation in the stationary Hamiltonian. Rotating atomic
quantum gases with such azimuthal gauge potentials differs from
those where metastable superflows are created by resonant pulses of
Raman LG beams~\cite{Ramanathan2011,Beattie2013} or mechanical
stirring~\cite{Wright2013}. In such cases the spinor wave function
is not the eigenstate of the Hamiltonian, therefore the gauge
potential is not well defined. Differing from applying Raman pulses
in the Rabi-flopping regime, one typically prepares dressed
eigenstates with gauge potentials by adiabatically sweeping the
detuning (see later text).

The light-induced azimuthal gauge potential can in principle be used
to measure superfluid fractions from the spin population
imbalance~\cite{Cooper2010}. Superfluid fraction corresponds to a
nonclassical rotational inertia
(NCRI)~\cite{Leggett1999,Leggett2001}, which is a property of
systems under thermal equilibrium, rather than of metastable
systems. In Refs.~\cite{Leggett1999,Leggett2001}, NCRI is manifested
in the Hess-Fairbank effect~\cite{Hess1967,Ishiguro2004}: with a
cylindrical container filled with $^4$He rotates at a sufficiently
low speed, after $^4$He is cooled below the superfluid transition
temperature, the atoms stop rotating and become out-of-equilibrium
with the container. NCRI is also an analog of the Meissner effect in
superconductors~\cite{Leggett1999,Leggett2001}.

Some attempts have been made to realize light-induced azimuthal
gauge potentials. In Ref.~\cite{Moulder2013}, Bose-Einstein
condensates (BECs) in a ring trap under LG Raman beams with the OAM
transfer between spin states ($|m_F\rangle\rightarrow
|m_F+1\rangle$) $\Delta\ell= 3\hbar$ are studied; a lifetime of
$\sim 0.1$~s of the lowest energy dressed state atoms is reported.
$F=1$ SOAMC BECs are demonstrated in Ref.~\cite{Chen2018} with the
gauge potential $A=0$, given that the atoms are in the middle-energy
dressed state in the $\langle \vec{F} \rangle=0$ polar phase. In
this paper, we present the first experimental realization of such
azimuthal gauge potential ($A\neq 0$) with SOAMC BECs in the lowest
energy dressed state.

\begin{figure*}
    \includegraphics[width=7 in]{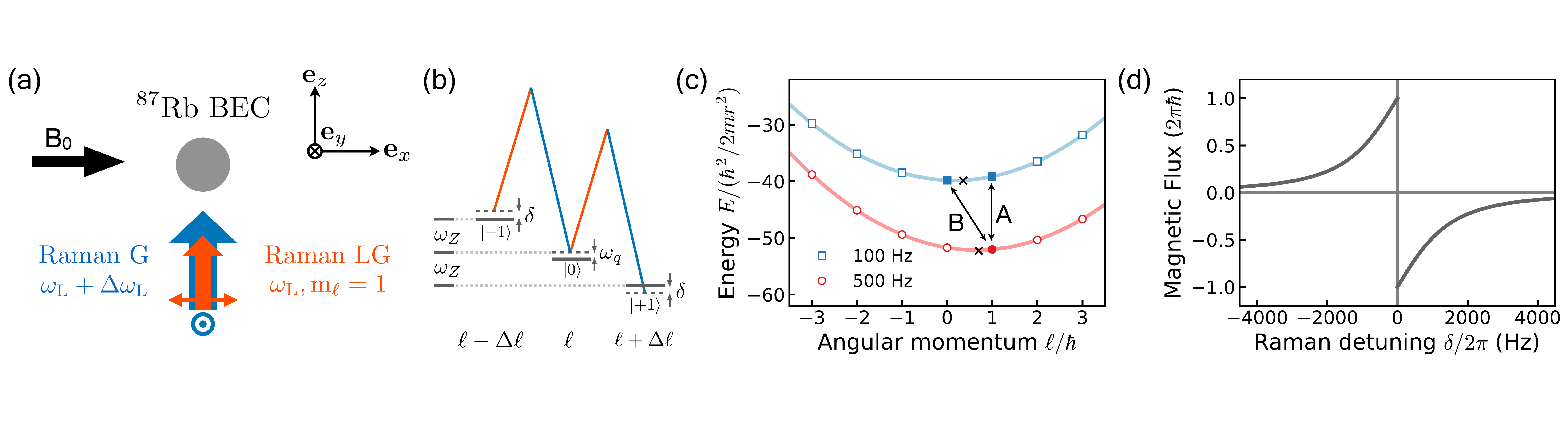}
    \caption{Experimental schematic and properties of the dressed state.
    (a) Experiment setup (b) Level diagram in $F=1$ manifold
    (c) Energy dispersion versus quasi-angular momentum $\ell$ at $r=2.0~\mu$m for
    $\delta/2\pi=100$ and 500 Hz. ``A" indicates $\ell=\hbar$ for both detunings
    in Fig.~2 data and ``B" indicates $\ell_g=0,\hbar$ for $\delta/2\pi=100$ and 500 Hz in
    Fig.~4,
    respectively. The x-marks indicate the minimum.
    (d) Synthetic magnetic flux $\Phi^{\pm}_{B^*}$ enclosed by the condensate radius $R$
    versus $\delta/2\pi$.}
\end{figure*}

SOAMC BECs can be implemented with an atom-light coupling
$\vec{\Omega}_{\rm eff}\cdot \vec{F}$ where the direction of
$\vec{\Omega}_{\rm eff}$ winds by a multiple of $2\pi$ as the
azimuthal angle $\phi$ increases from 0 to $2\pi$. Here,
$\vec{\Omega}_{\rm eff}$ is a light-induced effective magnetic
field~\cite{Goldman2013} typically realized by LG Raman
beams~\cite{Chen2018} and $\vec{F}$ is the atomic spin. Topological
excitations in spinor BECs, where the rich variety of order
parameters accommodates various types of topological
defects~\cite{Kawaguchi2012,Ueda2014}, can be created by versatile
design of $\vec{\Omega}_{\rm eff}$. This is analogous to the works
using spin rotation with real magnetic fields $\vec{B}$ with the
Hamiltonian term $\vec{B}\cdot \vec{F}$~\cite{Isoshima2000}, where
coreless vortices~\cite{Leanhardt2003},
monopoles~\cite{Ray2014,Ray2015,Hall2016,Ollikainen2017},
2D~\cite{Choi2012a,Choi2012} and 3D skyrmions~\cite{Lee2018}, and
the geometric Hall effect~\cite{Choi2013} are demonstrated.
Furthermore, 2D skyrmions~\cite{Leslie2009} and spin
monopoles~\cite{Hansen2016} with pulses of Raman LG beams are
achieved.

In this letter, we first load a BEC into the lowest energy
Raman-dressed state, creating a coreless
vortex~\cite{Kawaguchi2012}. Here one Raman beam is LG and carries
OAM. The dressed atoms experience azimuthal gauge potentials as
effective rotations, which we exploit to demonstrate the
Hess-Fairbank effect. As the synthetic magnetic flux arising from
the effective rotation is below a critical value, the dressed atoms
in the absolute ground state in the thermal equilibrium have zero
quasi-angular momentum $\ell^{\pm}$ (see later texts and
Eq.~\eqref{eq:gauge_transform}), being a coreless vortex. Above the
critical flux, a polar-core
vortex~\cite{Isoshima2001,Sadler2006,Kawaguchi2012} with nonzero
$\ell^{\pm}=\mp \hbar$ is achieved. We demonstrate the capability to
create both types of SO(3) vortices~\cite{Kawaguchi2012,Ueda2014},
the $\mathbb{Z}_2=0$ coreless vortex and $\mathbb{Z}_2=1$ polar-core
vortex (see the schematic drawing in Fig.~4b), in a unified and
controlled scheme. This opens a path for creating topological
excitations by tailoring atom-light interactions.

For atoms under sufficiently large $\vec{\Omega}_{\rm eff}\cdot
\vec{F}$ such that the motional kinetic energy $-(\hbar^2/2m)
\nabla^2$ is negligible, the energy eigenstates of the overall
Hamiltonian are well approximated by the eigenstates $|\xi_n\rangle$
of $\vec{\Omega}_{\rm eff}\cdot \vec{F}$. Under this approximation,
the atom's spinor wave function follows the local dressed eigenstate
$|\xi_n\rangle$, whose quantization axis is along $\vec{\Omega}_{\rm
eff}=\Omega(r)\cos \phi{\mathbf e}_{x}-\Omega(r)\sin \phi{\mathbf
e}_{y}+\delta{\mathbf e}_{z}$ given by our Raman beams. The state of
dressed atoms is
$\langle\vec{r}|\Psi\rangle=\varphi_n(\vec{r})|\xi_n(\vec{r})\rangle$,
where $\varphi_n$ is the external part and $|\xi_n\rangle$ is the
normalized spin part of the wave function. $|\varphi_n|=\sqrt{n_a}$
where $n_a$ is the density. The effective Hamiltonian for atoms
projected to $|\xi_n\rangle$, which governs the evolution of
$\varphi_n$, is~\cite{Chen2018}
\begin{eqnarray}
H_{\rm eff}^{(n)}=\frac{-\hbar^2}{2m}\nabla^2(r,z)
+\frac{\left(L_z-r A_{n}\right)^2}{2m r^2} + V(r)+\varepsilon_n+W_n.
\label{eq:projectedH}
\end{eqnarray}
Here $L_z= -i \hbar \partial_{\phi}$ is the angular momentum
operator for $\varphi_n$ with the eigenvalue $\ell$, and $A_{n}(r)=
(i\hbar/r)\langle \xi_n|\partial_{\phi} \xi_n\rangle$ is the
azimuthal gauge potential. $V(r)$ is the spin-independent trap,
$\varepsilon_n=n\sqrt{\Omega(r)^2+\delta^2}$ is the eigenenergy of
$\vec{\Omega}_{\rm eff}\cdot \vec{F}$, and $W_n\approx
\hbar^2/2mr^2$ is the geometric scalar potential. We label the
lowest, middle, and highest energy dressed states as
$|\xi_{-1}\rangle,|\xi_{0}\rangle,|\xi_{1}\rangle$, respectively.
$|\xi_{-1}\rangle$ is given by Euler rotations~\cite{Ho1998}
\begin{eqnarray}
|\xi_{-1}\rangle &= e^{i (\theta+\gamma)}\left( e^{i \phi}\frac{1 -
\cos \beta}{2}, -\frac{\sin \beta}{\sqrt{2}}, e^{-i \phi}\frac{1 +
\cos \beta}{2} \right)^{\rm T},
\end{eqnarray}
 where $\beta(r)=\tan^{-1} [ \Omega(r)/\delta]$ is the polar angle of
$\vec{\Omega}_{\rm eff}$, and $\theta+\gamma$ is the phase from a
gauge transformation. By choosing $\theta+\gamma=0$ for all
$\delta$, it leads to
\begin{eqnarray}\label{eq:gauge}
A_{-1}=(\hbar/r)\cos \beta,
\end{eqnarray}
where the angular momentum of $\varphi_{-1}$ is $\ell$ in this
gauge, and $\ell,\ell\pm\hbar$ are the mechanical angular momenta of
the bare spin $|m_F=0,\pm1\rangle$ components of the state
$\varphi_{-1}|\xi_{-1}\rangle$, respectively. In order to avoid a
singularity at $r=0$ in the synthetic magnetic field
$\vec{B}^{*}=\nabla\times\vec{A}$~\cite{Ho1996} as
$A_{-1}(r\rightarrow 0)\rightarrow \pm \hbar/r$ for $\delta
>(<)0$, we use alternative gauges with $\theta+\gamma=\pm \phi$, leading to
\begin{eqnarray}\label{eq:gauge_transform}
A_{-1}^{\pm}=\frac{\hbar}{r}(\cos \beta \mp 1), \ell^{\pm}=\ell\mp
\hbar
\end{eqnarray}
for $\delta>(<)0$, where $\ell^{\pm}$ is the angular momentum of the
external wave function $\varphi_{-1}^{\pm}$ in these gauges. Note
that the mechanical angular momentum
$\ell-rA_{-1}=\ell^{\pm}-rA_{-1}^{\pm}$ is gauge invariant.

\begin{figure*}
   \includegraphics[width=7 in]{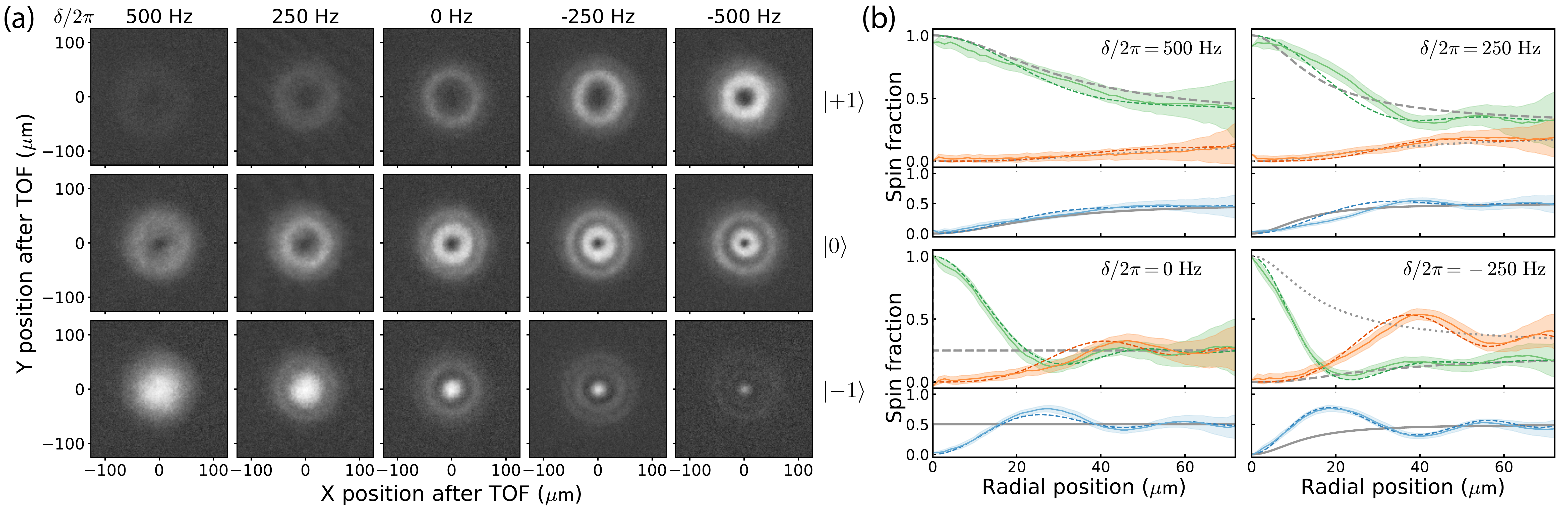}
    \caption{(a) Absorption images $D_{m_F}$ of the lowest energy
    dressed state projected onto bare spin $|m_F\rangle$
    states with various detuning $\delta$ after 24 ms TOF.
    (b) Experimental spin fractions (solid colored lines) versus the
    radial position are compared to the spin texture from $|\xi_{-1}\rangle$ (gray lines)
     and the TOF simulations from TDGPE (dashed colored lines). The green, orange
     and blue curves represents the
     $|-1\rangle,~|1\rangle,~|0\rangle$ components, respectively, where the
     shaded area indicates the standard deviation along ${\mathbf e}_{\phi}$. The
     in-situ spin textures Eq.~\eqref{eq:spin1} after a magnification of $r\rightarrow r/13.0$ after
     TOF~\cite{Castin1996,Chen2018}
     are shown as gray-dashed, -dotted, -solid lines for $|-1\rangle,~|1\rangle,~|0\rangle$ states.
    }
\end{figure*}

Consider atoms in the lowest energy dressed state $|\xi_{-1}\rangle$
prepared by loading a ground state BEC in $|m_F=-1\rangle$ with a
Thomas-Fermi (TF) wave function $\varphi_{\rm TF}=\sqrt{n_{\rm
TF}}$, where $\langle \vec{r}|\Psi\rangle_{t=0}= \varphi_{\rm
TF}\left(0,0,1\right)^{\rm T}$. The atoms are loaded to
$|\xi_{-1}\rangle$ as
\begin{eqnarray}\label{eq:spin1}
\langle \vec{r}|\Psi\rangle &= \varphi_{\rm TF}\left( e^{i
2\phi}\frac{1 - \cos \beta}{2}, -e^{i \phi}\frac{\sin
\beta}{\sqrt{2}},\frac{1 + \cos \beta}{2} \right)^{\rm T},
\end{eqnarray}
where the phase winding of the $m_F=-1$ component remains zero
during loading, and the potential $V(r)$ is cylindrically symmetric.
Eq.~\eqref{eq:spin1} has $\ell=\hbar$ using the gauge in
Eq.~\eqref{eq:gauge}, as shown in our data in Fig.~2. The atom's
spinor wave function follows $|\xi_{-1}\rangle$ only at $r\gtrsim
r_c$ due to the vanishing intensity of LG beam at $r = 0$, where
$r_c$ is the adiabatic radius. At $r\gtrsim r_c$, the radial kinetic
energy is negligible and for sufficiently slow $\dot{\delta}$
adiabatic loading is achieved.

We perform 3D time-dependent Gross-Pitaevskii equation (TDGPE)
calculations to simulate the loading of a BEC from $|m_F=-1\rangle$
into the dressed state $|\xi_{-1}\rangle$. We also solve the
absolute ground state using imaginary time propagations at given
$\Omega(r)$ and $\delta$.

Our experiment begins with $N\approx~ 1.2\times 10^5$ atoms in a
$^{87}$Rb BEC in the $|F=1, m_F=-1\rangle$ state in a crossed dipole
trap. The TF radius of the condensate is $R \approx 6.2~\mu$m along
${\mathbf e}_{r}$. With a bias magnetic field $B_0$ along ${\mathbf
e}_{x}$, the atoms experience a linear Zeeman shift
$\omega_Z/2\pi=0.57$~MHz and a quadratic Zeeman shift $\hbar
\omega_q \hat{F}_z^2$ with $\omega_q/2\pi= 50$ Hz. A Gaussian Raman
beam (G) and a Laguerre-Gaussian Raman beam (LG) with phase winding
$m_{\ell}=1$ co-propagate along ${\mathbf e}_{z}$, coupling atoms in
the $F=1$ manifold and transferring OAM of $\Delta\ell=\hbar$ to the
atoms (Fig.~1b).  The two Raman laser beams have wavelengths
$\lambda= 790$ nm with a frequency difference $\Delta \omega_L$ and
a Raman detuning $\delta=\Delta \omega_L- \omega_Z$ (Fig.~1ab). We
adiabatically load the BEC into the lowest energy dressed state
$|\xi_{-1}\rangle$ with final Raman coupling strength
$\Omega(r)=\Omega_M \sqrt{e}(r/r_M) e^{-r^2/2r_M^2}$. Here,
$\Omega_M/2\pi= 3.0$~kHz and $r_{M}=17~\mu$m is the peak intensity
radius.

The adiabatic loading is achieved by first turning on $\Omega(r)$ in
7~ms while holding the detuning at
$\delta/2\pi=\delta_f/2\pi+1.25$~kHz. Next, we sweep the detuning to
the final value $\delta_f$ in 7 ms where $-500~{\rm
Hz}<\delta_f/2\pi< 500~{\rm Hz}$. After preparing the atoms in the
dressed state and holding for a time $t_h$, we probe the atoms by
switching off the Raman beams and dipole trap simultaneously. After
a 24 ms time-of-flight (TOF) with all $|m_F\rangle$ components
expanding together, we take absorption images along ${\mathbf
e}_{z}$ for each $m_F$ states. Images with different final detuning
$\delta$ and $t_h=1$~ms are shown in Fig.~2a. For $\delta>0$, the
$|m_F=-1\rangle$ component has no hole, and $|0\rangle$ carries a
smaller hole than that of $|1\rangle$, consistent with the fact that
$|-1\rangle,|0\rangle,|1\rangle$ have angular momenta of $0,\hbar,
2\hbar$, respectively. The radial cross sections of the spin texture
$D_{m_F}/(D_1+D_0+D_{-1})$ are shown (Fig.~2b), which average over
the azimuthal angles. Here, $D_{m_F}$ is the optical density of
$|m_F\rangle$. We compare these spin textures with the local dressed
state $|\xi_{-1}\rangle$ in Eq.~\eqref{eq:spin1} and TOF simulations
from TDGPE.  The TDGPE simulation gives an $r_c \approx 1.8~\mu$m at
$\delta=0$, corresponding to the spin texture agreeing with
Eq.~\eqref{eq:spin1} at $r_{\rm TOF}\gtrsim 23~\mu$m.

We further study the stability of the lowest energy dressed state as
we hold the Raman field for a variable time $t_h$. With a deloading
procedure~\cite{Chen2018} which maps the dressed states
${|{\xi}_{-1}\rangle, |{\xi}_{0}\rangle, |{\xi}_{1}\rangle}$ back to
the bare spin states ${|-1\rangle, |0\rangle, |+1\rangle}$, we
measure the population in $|\xi_{n}\rangle$ versus $t_h$. We take
images along ${\mathbf e}_{y}$ after 14 ms-TOF with Stern-Gerlach
gradient at a variable $\delta$ with a $t_h$ up to 7.0~s. Fig.~3a
shows the atom number fraction in $|{\xi}_{n}\rangle$ over the total
number $|{\xi}_{-1}\rangle+|{\xi}_{0}\rangle+|{\xi}_{1}\rangle$ at
$\delta=0$ versus $t_h$.  The atoms slowly populate the excited
dressed states $|{\xi}_{0}\rangle$ and $|{\xi}_{1}\rangle$, and the
initial rates increase as the peak Raman coupling is reduced from
$\Omega_M/2\pi=3.0$~kHz to $1.5$~kHz. The lifetime of atom number
fraction in $|{\xi}_{-1}\rangle$ dropping to $50~\%$ is 4.5~s
(1.5~s) for $\Omega_M/2\pi=3.0~(1.5)$~kHz. Fig.~3b displays the atom
number fraction versus detuning at $t_h= 1$~s and
$\Omega_M/2\pi=3.0$~kHz, where the population transfer rates
decrease with increasing $|\delta|$. The results suggest the
population transfer rate into excited states increases with a
reduction of energy gap $\approx \sqrt{\Omega(r)^2+\delta^2}$, which
can be explained by the effects of thermal atoms. Our estimated
temperature $T\gtrsim 50$~nK~$\approx h \times 1$~kHz$/k_B$ is
comparable to the energy gap in the experiment.
\begin{figure}
    \includegraphics[width=3.4 in]{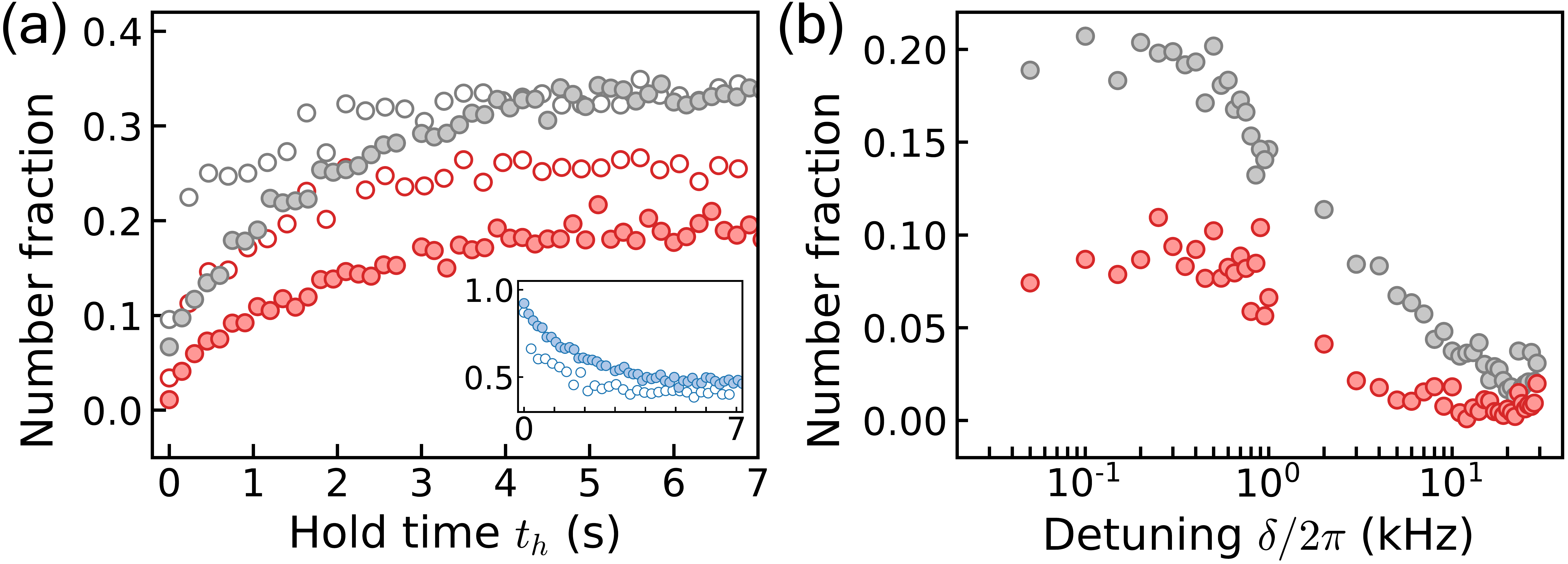}
    \caption{Stability of lowest energy dressed state $|\xi_{-1}\rangle$, shown as the number
    fraction in the excited dressed states
    $|\xi_0\rangle$ (gray symbols) and $|\xi_{1}\rangle$ (red symbols): (a) versus
    hold time $t_h$ at $\delta=0$ for $\Omega_M/2\pi=3.0$~kHz (solid symbols) and 1.5~kHz (open
    symbols). Inset shows the number fraction in $|\xi_{-1}\rangle$ versus $t_h$.
     (b) versus detuning with $t_h$=1~s and $\Omega_M/2\pi=3.0$~kHz.
     }
\end{figure}
Finally we demonstrate the Hess-Fairbank effect by studying the
absolute ground state of dressed atoms in the gauge fields. We begin
with thermal atoms right above the BEC transition temperature, load
to the lowest energy dressed state with detuning $\delta$, and
evaporatively cool the atoms to BEC. After a hold time $t_{h1}$ for
free evaporation, we probe the atoms after a 24~ms TOF. We note that
there is a different, but related experiment~\cite{Haljan2001},
where thermal bosonic atoms are evaporatively spun up and cooled to
reach condensation~\footnote{In Ref.~\cite{Haljan2001}, they study
vortex nucleation into a BEC in the environment of rotating thermal
atoms.}.

Similar to the Meissner effect, we tune the synthetic magnetic flux
via the detuning $\delta$ and observe the atoms in the absolute
ground state transit from having one quantum number to another. GPE
simulations show that the ground states are coreless vortices with
$\ell_g=\pm \hbar$ at $|\delta|/2\pi>210$~Hz and are polar-core
vortices with $\ell_g=0$ at $|\delta|/2\pi<210$~Hz (Fig.~4a);
$\ell_g$ is $\ell$ of the ground state. We obtain the averaged
absolute value of winding number $\langle|\ell_g|\rangle/\hbar$
versus detuning in Fig.~4, by taking TOF images of the bare spin
component $|m_F=0\rangle$, whose mechanical angular momentum is
$\ell_g$ under the gauge in Eq.~\eqref{eq:gauge}. At each $\delta$
we repeat the experiment for 20 times, observing $|m_F=0\rangle$ has
either no hole or a hole, indicating $\ell_g=0$ or $\ell_g=\pm
\hbar$, respectively. Such variations are due to the presence of a
root-mean-square (rms) detuning noise of $\approx h\times 70$~Hz (as
we repeat the experiment with the same $t_{h1}$, see supplement)
arising from the bias field noise in our setup. The curve
$\langle|\ell_g|\rangle$ vs. $\delta/2\pi$ is broadened and can
become slightly asymmetric at $\pm \delta$ since the detuning can
slowly vary during the time when the data is taken. A simulation
including a Gaussian-distributed detuning noise of $h \times 70$~Hz
rms is plotted in Fig.~4a, with a center shift of -70~Hz. The
observed ground state with small $|\delta|$ has
$\langle|\ell_g|\rangle/\hbar \sim 0$, corresponding to $\ell_g=0$
if there were no detuning noise. Fig.~4a shows the image of the
$\ell_g=0$ ($\ell_g=\hbar$) dressed state at
$\delta/2\pi=50~(400)$~Hz; the spin textures are schematically drawn
in Fig.~4b.

\begin{figure}
    \centering
    \includegraphics[width=3.4 in]{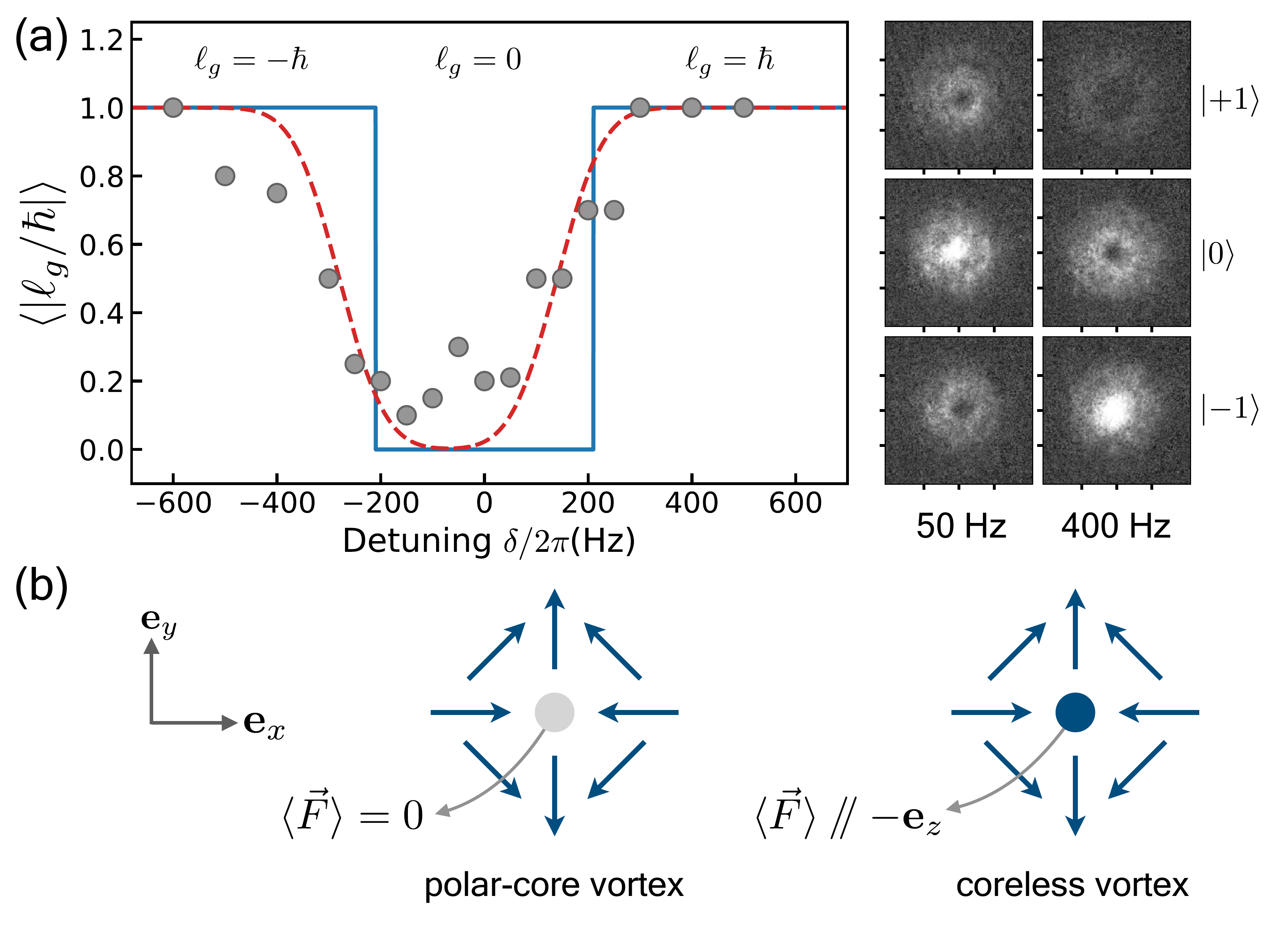}
    \caption{(a) Transition between $\ell=0$ and $\ell=\pm \hbar$
    states in the ground state.
    The measured average of magnitude of the quasi-angular momentum
    of the ground state atoms $\langle |\ell_g| \rangle$ versus detuning
    $\delta/2\pi$ (circles) with $t_{h1}\approx 0.2$~s. At small (large) $|\delta|$
     the ground state has $|\ell_g|=0$ ($\hbar$). Examples at $\delta/2\pi=50$ and 400~Hz are shown with
     the dressed states' bare spin components taken after TOF. The calculation for the ideal case
     (solid line) and a simulation including detuning noise in the experiment (dashed-line)
     are displayed; see text. (b) Schematic drawing of the
     spin textures at $\delta>0$, where the arrows show the direction of the transverse
     spin ($\langle F_x \rangle,\langle F_y\rangle$).}
\end{figure}

We compare our dressed atoms under gauge-induced rotations to BECs
under mechanical rotations~\cite{madison01,Aboshaeer01,Hodby2001}.
In Refs.~\cite{madison01,Aboshaeer01,Hodby2001}, Bose-condensation
is followed by mechanical stirring. The observed critical rotational
angular frequency for single-vortex nucleation is related to
dynamical instabilities, and is larger than the critical value
$\omega_c$ for the thermodynamic ground state to possess a single
vortex. By contrast, in our demonstration of the Hess-Fairbank
effect, the thermal atoms are subjected to the effective rotation
and then cooled to BEC in the thermodynamic ground state. Using the
gauges in Eq.~\eqref{eq:gauge_transform}, the synthetic magnetic
flux enclosed by the BEC's radius $R$ is $\Phi^{\pm }_{B^*}=h[\cos
\beta(R) \mp 1]$. Fig.~1d shows $|\Phi^{\pm}_{B^*}|$ approaches zero
at large $|\delta|$, and $\vec{B}^{*}$ is along $-(+){\mathbf
e}_{z}$ at $\delta>(<)~0$. Thus, our system is analogous to the
transition from zero to single-vortex ground state in rotating BECs
as the following: $\ell^{+}=0\rightarrow \ell^{+}=-\hbar$ (i.e.,
$\ell=\hbar\rightarrow \ell=0$) under $\Phi^{+}_{B^*}<0$. And
similarly for $\ell^{-}=0\rightarrow \ell^{-}=\hbar$ (i.e., $\ell=
-\hbar\rightarrow \ell=0$) under $\Phi^{-}_{B^*}>0$. At the
transition of $\delta/2\pi=\pm 210$~Hz, $|\Phi^{\pm}_{B^*}|= 0.8~h$
is smaller than the critical effective flux $\Phi_{\rm mech}$ for
mechanically rotating BECs. The flux is $\Phi_{\rm mech}= 2\pi
m\omega R^2=h(\omega/\omega_c)\ln(R/r_v)$ , where $\omega$ is the
angular frequency, $\omega_c=\hbar/(m
R^2)\ln(R/r_v)$~\cite{Lundh1997}, and $r_v$ is the vortex core size.
Thus the critical flux is $\Phi_{\rm mech}/h > 1$, while the
critical flux for our dressed state is $|\Phi^{\pm}_{B^*}|/h<1$.
This is due to the different form of $A(r)$ between our dressed
atoms and the mechanically rotating BECs (with symmetric gauge); see
supplement and a similar calculation in Ref.~\cite{Juzeliunas2005a}.

In conclusion, we demonstrate Raman-coupling-induced azimuthal gauge
potentials. They act as effective rotations which we exploit to show
Hess-Fairbank effect for the cold atoms. This work paves the way to
probe atomic superfluids with effective rotations, which can achieve
a stationary Hamiltonian and thermal equilibrium. This scheme
circumvents the issues from imperfect cylindrical symmetries in
mechanical rotations.

\begin{acknowledgments}
We thank W.~D. Phillips, C. Chin  and K.-Y. Lin for useful
discussions, and Y.-C.~Chen and J.-P. Wang for critical readings of
our manuscript. Y.~J.~L acknowledges the support of Career
Development Awards in Academia Sinica. S.~-K.~Y. was supported by
MOST. Y. K. was supported by JSPS KAKENHI JP15K17726. M.~-S.~C. and
M.~-J.~T. were supported by MOST 106-2112-M-001-033.

\emph{Note added}: After we completed the manuscript, we noticed a
recent work~\cite{Zhang2018} on pseudo-spin $1/2$ SOAMC BECs. They
reported observations similar to our demonstration of Hess-Fairbank
effect. Our work focuses on the perspective of effective rotations
enabled by light-induced azimuthal gauge potentials and
characterizing topological spin textures.

\end{acknowledgments}

\widetext
\clearpage
\begin{center}
\textbf{\large Supplemental Materials: Rotating atomic quantum gases with light-induced azimuthal gauge potentials and the observation of the Hess-Fairbank effect}
\end{center}
\setcounter{equation}{0}
\setcounter{figure}{0}
\setcounter{table}{0}
\setcounter{page}{1}
\makeatletter
\renewcommand{\theequation}{S\arabic{equation}}
\renewcommand{\thefigure}{S\arabic{figure}}
\renewcommand{\bibnumfmt}[1]{[S#1]}
\renewcommand{\citenumfont}[1]{S#1}

\maketitle

\section{TDGPE simulations}
We numerically simulate the dynamics by solving the three-component
3D time-dependent-Gross-Pitaevskii equation (TDGPE). This includes
the kinetic energies, quadratic Zeeman energy, mean field
interaction parameters $c_0=4\pi \hbar^2(a_0+2a_2)/3m$ and $c_2=4\pi
\hbar^2(a_2-a_0)/3m<0$, where $a_f$ is the s-wave scattering length
in the total spin $f$ channel~\cite{SHo1998}.

We use the Crank-Nicolson method and calculate in the system size of
$(256)^3$ grid points with grid size $0.16~\mu$m. During TOF, we
solve the full 3D TDGPE for up to $\le 4$~ms at which the
interatomic interaction energy becomes less than 3 percent of the
total energy. The further evolution is calculated by neglecting the
interaction term. The results for the lowest energy dressed state
with a short hold time $t_h=1$~ms are shown in Fig.~S1; our
corresponding data is in Fig.~2. Using the loaded atomic state from
TDGPE at $\delta=0$, the probability of projection to the local
dressed state $|\xi_{-1}\rangle$ is $\geq 0.99$ at $r\geq
r_c=1.8~\mu$m.

\section{Synthetic magnetic flux of dressed atoms}
We compare the thermodynamic ground state of the dressed atoms in
$|\xi_{-1}\rangle$ under synthetic magnetic fields to that of
mechanically rotating BECs. Taking $\delta>0$ where
$\vec{B}^{*}\cdot {\mathbf e}_{z}<0$ as the example, we consider the
azimuthal kinetic energy in Eq.~(1), which is gauge invariant and
can be written as $(\ell^{+}-rA_{-1}^{+})^2/2mr^2$ in the gauge of
Eq.~(4). Here we simplify the calculation by assuming a disk BEC
geometry, although our BEC is 3D and has a Thomas-Fermi (TF)
profile. Similar calculations for 3D TF BECs are shown in
Ref.~\cite{SLundh1997}. The contribution to the kinetic energy per
atom is
\begin{align}\label{eq:Energy}
&\frac{1}{\pi R^2}\int_{0}^{R} dr 2\pi
r\frac{(\ell^{+}-rA_{-1}^{+})^2}{2mr^2}\nonumber\\
&=\frac{1}{\pi R^2}\left[\int dr 2\pi r
\frac{\left(\ell^{+}\right)^2}{2mr^2}-\int dr 2\pi r \frac{\ell^{+}
\hbar}{mr^2}(\cos \beta -1)+\int dr 2\pi
r\frac{\left(rA_{-1}^{+}\right)^2}{2mr^2}\right].
\end{align}
With $\delta>0$, $\cos \beta -1$ is negative. We thus compare the
energy with $\ell^{+}=0$ and with $\ell^{+}=-\hbar$, where the lower
one is the ground state. The energy for $\ell^{+}=-\hbar$ relative
to $\ell^{+}=0$ is
\begin{align}\label{eq:Edressed}
E_{B^{*}}&=\frac{1}{\pi R^2}\int_{r_{v}}^{R} dr 2\pi r
\frac{\hbar^2}{2mr^2}+\frac{1}{\pi R^2}\int_{0}^{R} dr 2\pi r
\frac{\hbar^2}{mr^2}(\cos \beta -1),
\end{align}
where the first term is the vortex kinetic energy $E_v$ and $r_{v}$
is the vortex core size. At large $\delta$, $\cos \beta -1\approx 0$
and $E_{B^{*}}\approx E_v$, leading to the $\ell^+=0$ ground state.
As $\delta$ decreases, $|\cos \beta -1|$ increases, and the ground
state makes a transition to $\ell^+=-\hbar$ when the absolute value
of the second therm in Eq.~\eqref{eq:Edressed} equals $E_v$.

\begin{figure}
    \centering
    \includegraphics[width=7.0 in]{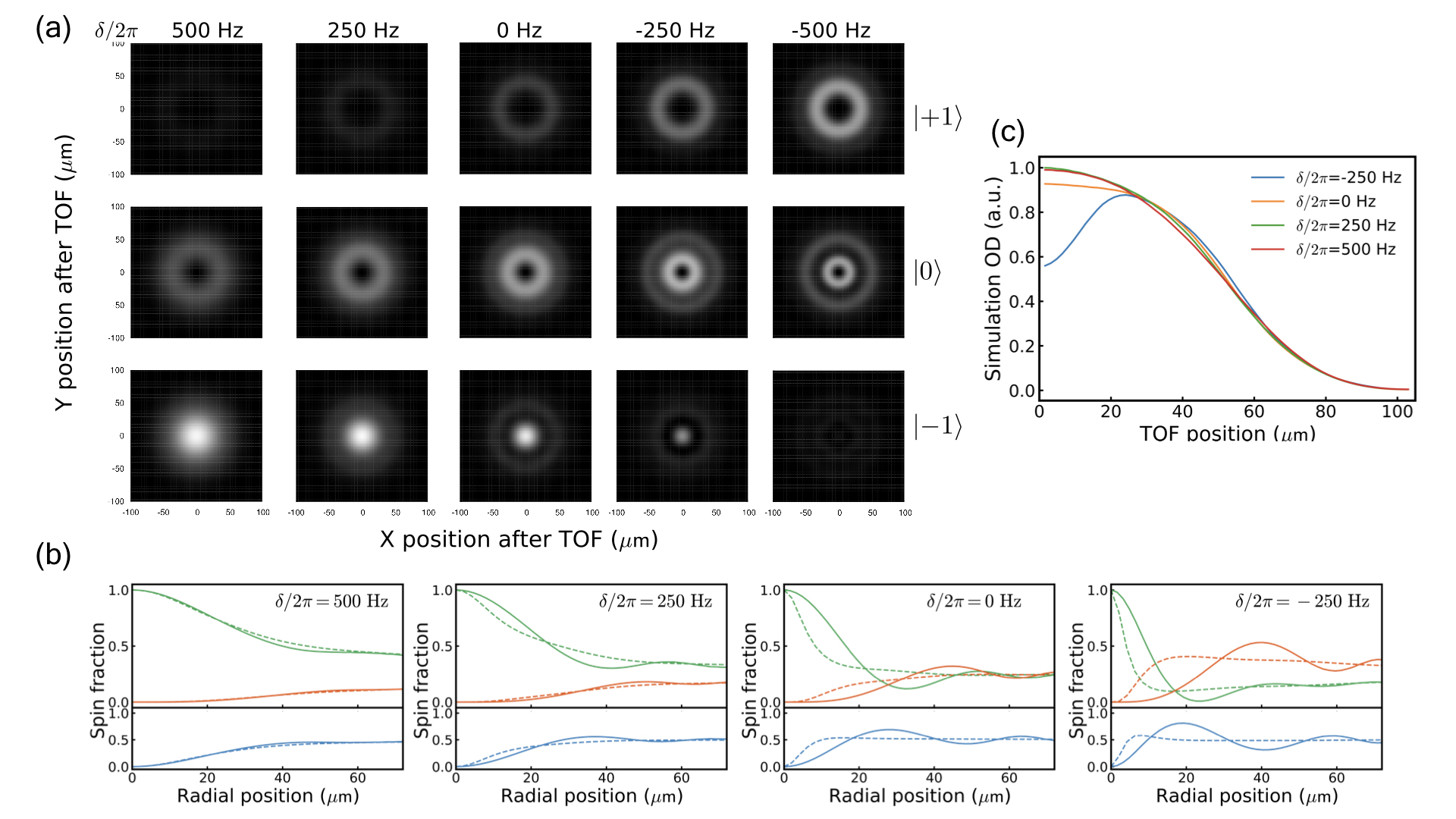}
    \caption{3D TDGPE simulation results for the BEC loaded into the $\ell=\hbar$
    lowest energy dressed state at various detuning $\delta$ with a hold time $t_h=1$~ms.
    (a) Absorption images $D_{m_F}$ after a 24~ms
    TOF. The image field of view is 200$\times$200~$\mu$m.
    (b) Spin texture $D_{m_F}/(D_1+D_0+D_{-1})$ vs. radial position. Dashed curves denote the
    in-situ profile
    after a $r\rightarrow r/13.0$ magnification; solid curves denote
    those after a 24 ms TOF. Green, orange and blue
    curves denote $|-1\rangle,|1\rangle,|0\rangle$, respectively.
    (c) total optical density $(D_1+D_0+D_{-1})$ for
    simulated TOF profiles.
    }
\end{figure}

Then we analogously consider a mechanically rotating BEC with
angular frequency $-\omega$, where $\omega>0$ . The kinetic energy
for $\ell=-\hbar$ relative to $\ell=0$ is
\begin{align}\label{eq:Emech}
E_{\rm mech}&=\frac{1}{\pi R^2}\int_{r_{v}}^{R} dr 2\pi r
\frac{\hbar^2}{2mr^2}-\frac{1}{\pi R^2}\int_{0}^{R} dr 2\pi r
\frac{\hbar^2}{mr^2}\frac{m\omega r^2}{\hbar}\nonumber\\
&=E_v-\frac{1}{\pi R^2}\int_{0}^{R} dr 2\pi r \hbar \omega.
\end{align}
The critical angular frequency for the ground state changes from
$\ell^+=0$ to $\ell^{+}=-\hbar$ with increasing $\omega$ is
$\omega_c=\hbar/(m R^2)\ln(R/r_{v})$.

\begin{figure}
    \centering
    \includegraphics[width=6.5 in]{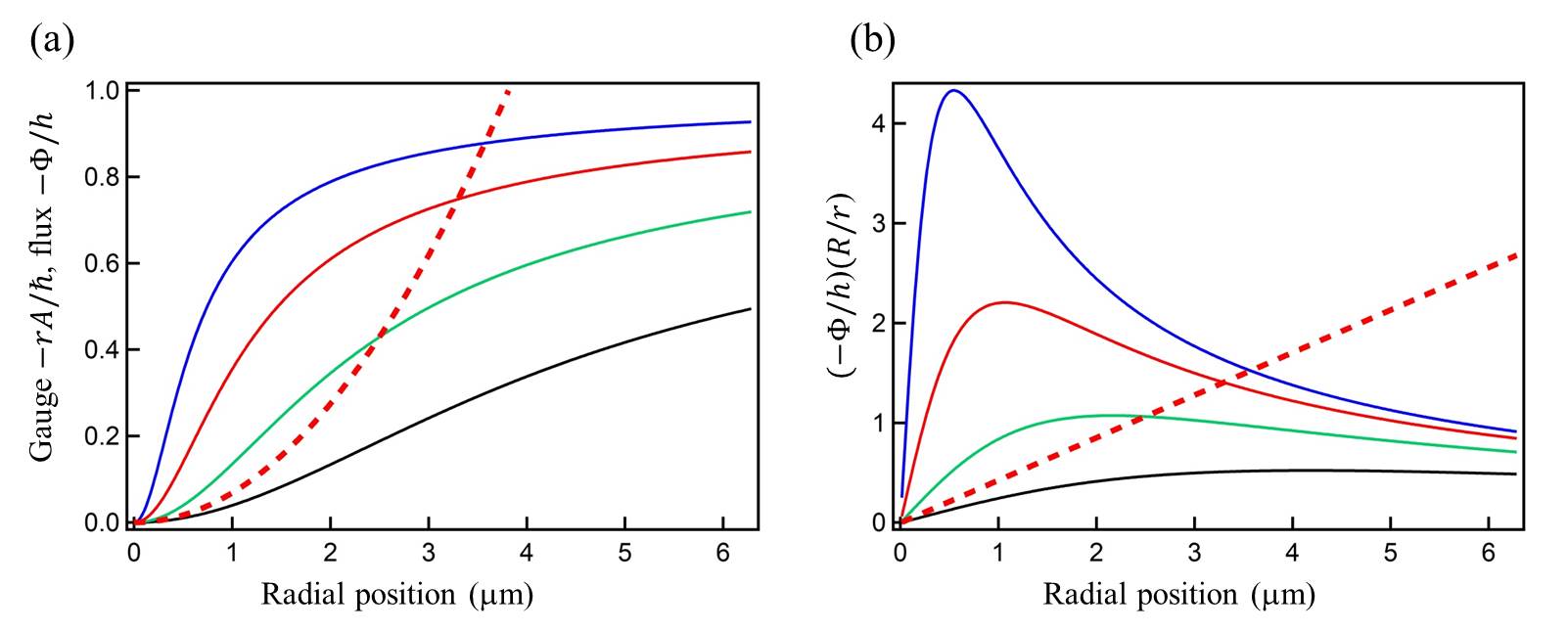}
    \caption{(a) Gauge potential $-rA_{-1}^{+}/\hbar$, which is equal to the flux $-\Phi/h$,
    versus radial position $r$. Black, green, red and blue solid curves denote the
    lowest energy dressed state with $\delta/2\pi= 1000,500,295,125$~Hz, respectively.
    $\delta_c/2\pi= 295$~Hz is the critical detuning. Red dashed
    curve represents the mechanically rotating BEC with the critical frequency $\omega_c$.
    (b) $-(\Phi/h)(R/r)$ versus $r$. Symbols represent the same as those in (a).}
\end{figure}

To compare the dressed atoms and the mechanically rotating BEC, we
consider the dimensionless gauge potential $rA/\hbar$, which is
equal to the flux $\Phi$ enclosed by radius $r$ in unit of $h$, given that $\oint \vec{A}\cdot\vec{dl}=A(r)2\pi r=\Phi$;
$\Phi$ represents $\Phi^{+}_{B^*}$ and $\Phi_{\rm mech}$. We then
plot $rA_{-1}^{+}/\hbar=\cos \beta(r) -1$ and $m\omega_c r^2/\hbar$
versus $r$ in Fig.~S2a, showing curves of $\cos \beta(r) -1$ with
various detuning $\delta/2\pi$ for the dressed atoms. Here we use
$r_{v}=0.5~\mu$m for both Eq.~\eqref{eq:Edressed} and
Eq.~\eqref{eq:Emech}, and thus $\omega_c/2\pi=8.0$~Hz for the TF
radius $R\approx 6~\mu$m. We derive $r_v$ for dressed atoms from the
projection probability of the spinor wave function of the absolute
ground state with $\ell^+=-\hbar$ (i.e., $\ell=0$) onto
$|\xi_{-1}\rangle$, where Eq.~\eqref{eq:Energy} applies. This
probability is zero at $r=0$ and increases with increasing $r$, and
we take $r_v$ as the radial position with the probability of 0.65.
We find $r_v~\sim~0.5~\mu$m is approximately independent of $\delta$
for 125~Hz$<\delta/2\pi<$~300~Hz.

From Fig.~S2a, we observe the flux enclosed by $r=R$ is $|\Phi_{\rm
mech}|/h\approx 2.6> 1$ while $|\Phi^{+}_{B^*}|/h<1$ is limited by
our $\Delta \ell=\hbar$. Since $E_v$ is the same in
Eq.~\eqref{eq:Edressed} and Eq.~\eqref{eq:Emech}, we then compare
the integrand of the second term in Eq.~\eqref{eq:Edressed} and
Eq.~\eqref{eq:Emech}, respectively, versus $r$ in Fig.~S2b, which
are contributed from the gauge potential $rA_{-1}^{+}$ and
$m\omega_c r^2$. The integrand is scaled as $(\Phi/h)(R/r)$. At the
critical detuning $\delta_c$ for the dressed atoms, the
quasi-angular momentum $\ell^{+}=\ell-\hbar$ of the ground state
changes from 0 to $-\hbar$. We determine $\delta_c$ from Fig.~S2b,
where the enclosed area by the red curve denoting $\delta_c$ equals
to that of the red-dashed curve representing the mechanically
rotating BEC with the critical frequency $\omega_c$. We find this
condition is fulfilled at $\delta_c/2\pi\approx 295$~Hz. Fig.~S2a
shows that at small $r$ the gauge potential $|rA_{-1}|$ of dressed
atoms with $\delta/2\pi\leq 500$~Hz is larger than the $m\omega_c
r^2$ for the mechanical rotation; the contribution at small $r$ is
further enhanced by the $2\pi \hbar^2/(mr)$ factor in the integrand
(see Fig.~S2b). Therefore, the critical flux for the symmetric gauge
is $|\Phi_{\rm mech}|/h> 1$ while $|\Phi^{+}_{B^*}|/h\approx 0.8<1$
for our dressed atoms.

\section{Experimental setup and procedures}
For the coreless vortex data in Fig.~2, we first produce a $^{87}$Rb
BEC and then load it into the Raman-dressed state
$|\xi_{-1}\rangle$. We achieve a BEC of $N \approx 1.2 \times 10^5$
atoms in a crossed dipole trap in
$|F,m_F\rangle=|1,-1\rangle$~\cite{SChen2018}. The dipole trap
contains two 1064 nm laser beams propagating along ${\mathbf
e}_{x^{'}},{\mathbf e}_{y^{'}}=({\mathbf e}_{x}\pm{\mathbf
e}_{y})/\sqrt{2}$ with beam waists of $\sim 30~\mu$m. After the
forced evaporation and a 1.5 s free evaporation, the dipole beam
powers are ramped up in 0.2 s in order to reach the final trap
frequencies of
$(\omega_{x^{'}},\omega_{y^{'}},\omega_z)/2\pi$=(140,140,190)~Hz.
The radial trap frequency is chosen such that it is sufficient for
the atoms to sustain a Thomas-Fermi profile in the presence of the
anti-trapping light shift potential near Raman resonance,
$-\sqrt{\Omega(r)^2+\delta^2}$.

\begin{figure}
    \centering
    \includegraphics[width=7.0 in]{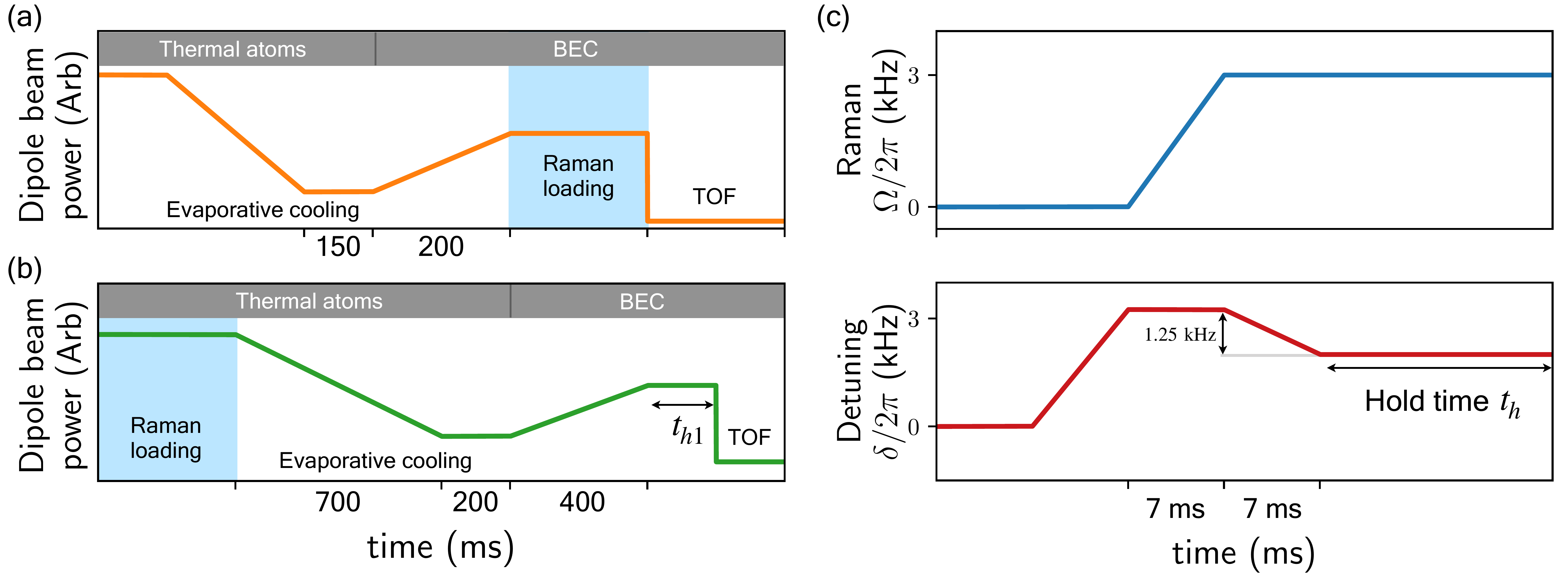}
    \caption{Time sequences of the experiment. (a) Dipole beam power in the procedure
    for the coreless vortex data in Fig.~2. (b) Dipole beam power in the procedure
    for data in Fig.~4 showing Hess-Fairbank effect. (c) The sequences of Raman loading in (a) and (b).}
\end{figure}

After the BEC preparation we wait for the external trigger from the
60 Hz line, after which we apply feed-forward current signals into
bias coils to cancel the field noise from 60 Hz harmonics (see later
discussions). Then we load the $|m_F=-1\rangle$ BEC into the lowest
energy dressed state $|\xi_{-1}\rangle$ with the following
procedures. We ramp the detuning to
$\delta/2\pi=\delta_f/2\pi+1.25$~kHz while the Raman beams are off,
ramp $\Omega(r,t)$ in 7~ms to the final value of $\Omega_M/2\pi=
3.0$ kHz, and then ramp the detuning to $\delta_f/2\pi$ between 500
Hz and -500 Hz in 7 ms (see Fig.~S3), subsequently holding
$\Omega_M$ and $\delta$ at constant for $t_h$. The Raman beams are
at $\lambda=790$~nm where their scalar light shifts from the D1 and
D2 lines cancel. The Gaussian Raman beam has a waist of $200~\mu$m,
and the LG Raman beam produced by a vortex phase plate has a phase
winding number $m_{\ell}=1$ and radial index of 0. The Raman beams
are linearly polarized along ${\mathbf e}_{x}$ and ${\mathbf
e}_{y}$, respectively.

To show the Hess-Fairbank effect in Fig.~4, we start with thermal
atoms right above the BEC transition temperature, load to the
dressed state $|\xi_{-1}\rangle$ using the same sequence as that for
data in Fig.~2. Next we force evaporatively cool the atoms to
achieve BECs and hold 0.2~s for free evaporation, then increase the
dipole beam powers in 0.4~s to reach the trap frequencies
$(\omega_{x^{'}},\omega_{y^{'}},\omega_z)$, subsequently holding for
$t_{h1}$ before TOF. To reduce the shot-to-shot field noise when TOF
starts, we wait for the 60 Hz line trigger at 20~ms before the TOF.
This adds to $t_{h1}$ by a varying time up to 16.7~ms (period of 60
Hz) as the experiment is repeated.

For projection measurements of the spinor state $|\xi_s\rangle$, we
abruptly turn off the dipole trap and Raman beams, simultaneously
and adiabatically rotate the magnetic bias field from along
${\mathbf e}_{x}$ to that along the imaging beam direction within
$0.4$ ms. The spinor wave function is then projected to the bare
spin $m_F$ basis. For data in Fig.~2 and Fig.~4, the atoms then
expand in free space with all $m_F$ components together for a TOF.
To perform spin-selective imaging, we apply microwave spectroscopy
for imaging each $|m_F\rangle$~\cite{SChen2018}. Each $|m_F\rangle$
image is obtained in an individual experimental realization. After
the $F=1$ atoms are transferred to $F=2$ by the microwave pulses, we
apply a resonant absorption imaging pulse of $\sim 14~\mu$s with
$\sigma +$ polarization at the $|F=2,m_F=2\rangle \rightarrow
|F^{'}=3,m_F=3\rangle$ cycling transition. With our $I/I_{\rm sat}$
parameters, we use the modified Beer-Lamberet
law~\cite{SReinaudi2007} to derive correct optical densities. For
data in Fig.~3, we apply a Stern-Gerlach field gradient to spatially
separate individual $|m_F\rangle$ states during TOF, then a
$F=1\rightarrow F^{'}=2$ repumping pulse is applied, after which the
images of all $|m_F\rangle$ states are taken in a single shot.

We characterize the ambient field noise after the external trigger
from the 60 Hz line. The noise is dominated by the 60 Hz line signal
with a standard deviation ($\sigma$) of $\sim h\times 300$~Hz. After
we apply feed-forward signals in the bias fields to cancel the
dominating field noise at 60 Hz, the $1-\sigma$ residual field noise
can be reduced to $h \times 120$~Hz within 0.4~s after the 60 Hz
trigger, in the best case. For longer time the 60~Hz line has a
phase decoherence and the feed-forward cancelation does not work.
For the data in Fig.~2 where the Raman loading takes 14~ms, we
prepare the dressed state after the 60 Hz line trigger in order to
reduce the shot-to-shot field variation with a fixed time after the
trigger, and the measured $1-\sigma$ field noise from repeated
experimental shots is $ \sim h\times 70$~Hz.

\section{Data analysis}
For data in Fig.~2, we average over 4 to 5 images for each $\delta$
taken under identical conditions. We post-select images whose vortex
positions in $|m_F=\pm 1\rangle$ with respect to the BEC center are
$<0.3~\mu$m (converted from TOF position to the in-situ position).
We determine $\delta/2\pi$ from the rf-spectroscopy with an
uncertainty of $\approx 20$~Hz. The uncertainty of the spin fraction
displayed in Fig.~2b is $\sigma/\sqrt{N}$, where $\sigma$ is the
standard deviation of pixels along ${\mathbf e}_{\phi}$ at a fixed
$r$, and $N=4$ or 5 is the number of images.

For data in Fig.~4, the variations in the winding number of
$|m_F=0\rangle$ component within 20 identical experimental
realizations are most likely due to the detuning noise. Thus, in the
images at $\delta/2\pi=50$~Hz with $\langle|\ell_g|\rangle/\hbar
\sim 0$, for $|m_F=0\rangle$ component we post-select those with no
hole corresponding to $\ell_g=0$. For $|m_F=1\rangle$ component we
select those with smaller holes corresponding to
$\ell_g+\hbar=\hbar$ and excluding those with larger holes
corresponding to $\ell_g+\hbar=2\hbar$. Similarly, we post-select
those with $\ell_g-\hbar=-\hbar$ in $|m_F=-1\rangle$ with
$\delta/2\pi=50$~Hz, and further select those with $\ell_g=\hbar$
for $|m_F=0,\pm 1\rangle$ images with $\delta/2\pi=400$~Hz based on
the hole size in the images. For both $\delta/2\pi=50,400$~Hz, each
$m_F$ state image is a single-shot image.

In the Hess-Fairbank effect experiment, we consider the effects of
the dominating detuning noise at 60~Hz with an amplitude of $\approx
h\times 420$~Hz. After BEC is reached, the sum of the hold time and
the ramp time of dipole beam power is 0.8~s. After the 60 Hz line
signal decoheres, the feed-forward cancelation does not function,
and thus the 60~Hz detuning noise could drive the atoms slightly out
of the equilibrium and out of the absolute ground state. Further
reduction of bias field noise is needed in order to improve the
current measurements.


\begin{thebibliography}{53}
\expandafter\ifx\csname
natexlab\endcsname\relax\def\natexlab#1{#1}\fi
\expandafter\ifx\csname bibnamefont\endcsname\relax
  \def\bibnamefont#1{#1}\fi
\expandafter\ifx\csname bibfnamefont\endcsname\relax
  \def\bibfnamefont#1{#1}\fi
\expandafter\ifx\csname citenamefont\endcsname\relax
  \def\citenamefont#1{#1}\fi
\expandafter\ifx\csname url\endcsname\relax
  \def\url#1{\texttt{#1}}\fi
\expandafter\ifx\csname urlprefix\endcsname\relax\def\urlprefix{URL
}\fi \providecommand{\bibinfo}[2]{#2}
\providecommand{\eprint}[2][]{\url{#2}}

\bibitem[{\citenamefont{Lin et~al.}(2009)\citenamefont{Lin, Compton, Perry,
  Phillips, Porto, and Spielman}}]{lin09}
\bibinfo{author}{\bibfnamefont{Y.-J.} \bibnamefont{Lin}},
  \bibinfo{author}{\bibfnamefont{R.~L.} \bibnamefont{Compton}},
  \bibinfo{author}{\bibfnamefont{A.~R.} \bibnamefont{Perry}},
  \bibinfo{author}{\bibfnamefont{W.~D.} \bibnamefont{Phillips}},
  \bibinfo{author}{\bibfnamefont{J.~V.} \bibnamefont{Porto}}, \bibnamefont{and}
  \bibinfo{author}{\bibfnamefont{I.~B.} \bibnamefont{Spielman}},
  \bibinfo{journal}{Phys. Rev. Lett.} \textbf{\bibinfo{volume}{102}},
  \bibinfo{eid}{130401} (\bibinfo{year}{2009}).

\bibitem[{\citenamefont{Aidelsburger et~al.}(2013)\citenamefont{Aidelsburger,
  Atala, Lohse, Barreiro, Paredes, and Bloch}}]{aidelsburger2013}
\bibinfo{author}{\bibfnamefont{M.}~\bibnamefont{Aidelsburger}},
  \bibinfo{author}{\bibfnamefont{M.}~\bibnamefont{Atala}},
  \bibinfo{author}{\bibfnamefont{M.}~\bibnamefont{Lohse}},
  \bibinfo{author}{\bibfnamefont{J.~T.} \bibnamefont{Barreiro}},
  \bibinfo{author}{\bibfnamefont{B.}~\bibnamefont{Paredes}}, \bibnamefont{and}
  \bibinfo{author}{\bibfnamefont{I.}~\bibnamefont{Bloch}},
  \bibinfo{journal}{Phys. Rev. Lett.} \textbf{\bibinfo{volume}{111}},
  \bibinfo{pages}{185301} (\bibinfo{year}{2013}).

\bibitem[{\citenamefont{Miyake et~al.}(2013)\citenamefont{Miyake, Siviloglou,
  Kennedy, Burton, and Ketterle}}]{miyake2013}
\bibinfo{author}{\bibfnamefont{H.}~\bibnamefont{Miyake}},
  \bibinfo{author}{\bibfnamefont{G.~A.} \bibnamefont{Siviloglou}},
  \bibinfo{author}{\bibfnamefont{C.~J.} \bibnamefont{Kennedy}},
  \bibinfo{author}{\bibfnamefont{W.~C.} \bibnamefont{Burton}},
  \bibnamefont{and} \bibinfo{author}{\bibfnamefont{W.}~\bibnamefont{Ketterle}},
  \bibinfo{journal}{Phys. Rev. Lett.} \textbf{\bibinfo{volume}{111}},
  \bibinfo{pages}{185302} (\bibinfo{year}{2013}).

\bibitem[{\citenamefont{Struck et~al.}(2012)\citenamefont{Struck,
  {\"O}lschl{\"a}ger, Weinberg, Hauke, Simonet, Eckardt, Lewenstein, Sengstock,
  and Windpassinger}}]{Struck2012}
\bibinfo{author}{\bibfnamefont{J.}~\bibnamefont{Struck}},
  \bibinfo{author}{\bibfnamefont{C.}~\bibnamefont{{\"O}lschl{\"a}ger}},
  \bibinfo{author}{\bibfnamefont{M.}~\bibnamefont{Weinberg}},
  \bibinfo{author}{\bibfnamefont{P.}~\bibnamefont{Hauke}},
  \bibinfo{author}{\bibfnamefont{J.}~\bibnamefont{Simonet}},
  \bibinfo{author}{\bibfnamefont{A.}~\bibnamefont{Eckardt}},
  \bibinfo{author}{\bibfnamefont{M.}~\bibnamefont{Lewenstein}},
  \bibinfo{author}{\bibfnamefont{K.}~\bibnamefont{Sengstock}},
  \bibnamefont{and}
  \bibinfo{author}{\bibfnamefont{P.}~\bibnamefont{Windpassinger}},
  \bibinfo{journal}{Phys. Rev. Lett.} \textbf{\bibinfo{volume}{108}},
  \bibinfo{pages}{225304} (\bibinfo{year}{2012}).

\bibitem[{\citenamefont{Parker et~al.}(2013)\citenamefont{Parker, Ha, and
  Chin}}]{Parker2013}
\bibinfo{author}{\bibfnamefont{C.~V.} \bibnamefont{Parker}},
  \bibinfo{author}{\bibfnamefont{L.-C.} \bibnamefont{Ha}}, \bibnamefont{and}
  \bibinfo{author}{\bibfnamefont{C.}~\bibnamefont{Chin}},
  \bibinfo{journal}{Nature Physics} \textbf{\bibinfo{volume}{9}},
  \bibinfo{pages}{769} (\bibinfo{year}{2013}).

\bibitem[{\citenamefont{Jotzu et~al.}(2014)\citenamefont{Jotzu, Messer,
  Desbuquois, Lebrat, Uehlinger, Greif, and Esslinger}}]{jotzu2014}
\bibinfo{author}{\bibfnamefont{G.}~\bibnamefont{Jotzu}},
  \bibinfo{author}{\bibfnamefont{M.}~\bibnamefont{Messer}},
  \bibinfo{author}{\bibfnamefont{R.}~\bibnamefont{Desbuquois}},
  \bibinfo{author}{\bibfnamefont{M.}~\bibnamefont{Lebrat}},
  \bibinfo{author}{\bibfnamefont{T.}~\bibnamefont{Uehlinger}},
  \bibinfo{author}{\bibfnamefont{D.}~\bibnamefont{Greif}}, \bibnamefont{and}
  \bibinfo{author}{\bibfnamefont{T.}~\bibnamefont{Esslinger}},
  \bibinfo{journal}{Nature} \textbf{\bibinfo{volume}{515}},
  \bibinfo{pages}{237} (\bibinfo{year}{2014}).

\bibitem[{\citenamefont{Dalibard et~al.}(2011)\citenamefont{Dalibard, Gerbier,
  Juzeli\ifmmode~\bar{u}\else \={u}\fi{}nas, and \"Ohberg}}]{Dalibard11}
\bibinfo{author}{\bibfnamefont{J.}~\bibnamefont{Dalibard}},
  \bibinfo{author}{\bibfnamefont{F.}~\bibnamefont{Gerbier}},
  \bibinfo{author}{\bibfnamefont{G.}~\bibnamefont{Juzeli\ifmmode~\bar{u}\else
  \={u}\fi{}nas}}, \bibnamefont{and}
  \bibinfo{author}{\bibfnamefont{P.}~\bibnamefont{\"Ohberg}},
  \bibinfo{journal}{Rev. Mod. Phys.} \textbf{\bibinfo{volume}{83}},
  \bibinfo{pages}{1523} (\bibinfo{year}{2011}).

\bibitem[{\citenamefont{Goldman et~al.}(2014)\citenamefont{Goldman, Juzeliunas,
  \"Ohberg, and Spielman}}]{Goldman2013}
\bibinfo{author}{\bibfnamefont{N.}~\bibnamefont{Goldman}},
  \bibinfo{author}{\bibfnamefont{G.}~\bibnamefont{Juzeliunas}},
  \bibinfo{author}{\bibfnamefont{P.}~\bibnamefont{\"Ohberg}}, \bibnamefont{and}
  \bibinfo{author}{\bibfnamefont{I.~B.} \bibnamefont{Spielman}},
  \bibinfo{journal}{Rep. Prog. Phys.} \textbf{\bibinfo{volume}{77}},
  \bibinfo{pages}{126401} (\bibinfo{year}{2014}).

\bibitem[{\citenamefont{Zhai}(2015)}]{Zhai2015}
\bibinfo{author}{\bibfnamefont{H.}~\bibnamefont{Zhai}},
  \bibinfo{journal}{Reports on Progress in Physics}
  \textbf{\bibinfo{volume}{78}}, \bibinfo{pages}{026001}
  (\bibinfo{year}{2015}).

\bibitem[{\citenamefont{Lin et~al.}(2011)\citenamefont{Lin, Jimenez-Garcia, and
  Spielman}}]{Lin11}
\bibinfo{author}{\bibfnamefont{Y.~J.} \bibnamefont{Lin}},
  \bibinfo{author}{\bibfnamefont{K.}~\bibnamefont{Jimenez-Garcia}},
  \bibnamefont{and} \bibinfo{author}{\bibfnamefont{I.~B.}
  \bibnamefont{Spielman}}, \bibinfo{journal}{Nature}
  \textbf{\bibinfo{volume}{471}}, \bibinfo{pages}{83} (\bibinfo{year}{2011}).

\bibitem[{\citenamefont{Wu et~al.}(2016)\citenamefont{Wu, Zhang, Sun, Xu, Wang,
  Ji, Deng, Chen, Liu, and Pan}}]{Wu2016}
\bibinfo{author}{\bibfnamefont{Z.}~\bibnamefont{Wu}},
  \bibinfo{author}{\bibfnamefont{L.}~\bibnamefont{Zhang}},
  \bibinfo{author}{\bibfnamefont{W.}~\bibnamefont{Sun}},
  \bibinfo{author}{\bibfnamefont{X.-T.} \bibnamefont{Xu}},
  \bibinfo{author}{\bibfnamefont{B.-Z.} \bibnamefont{Wang}},
  \bibinfo{author}{\bibfnamefont{S.-C.} \bibnamefont{Ji}},
  \bibinfo{author}{\bibfnamefont{Y.}~\bibnamefont{Deng}},
  \bibinfo{author}{\bibfnamefont{S.}~\bibnamefont{Chen}},
  \bibinfo{author}{\bibfnamefont{X.-J.} \bibnamefont{Liu}}, \bibnamefont{and}
  \bibinfo{author}{\bibfnamefont{J.-W.} \bibnamefont{Pan}},
  \bibinfo{journal}{Science} \textbf{\bibinfo{volume}{354}},
  \bibinfo{pages}{83} (\bibinfo{year}{2016}).

\bibitem[{\citenamefont{Huang et~al.}(2016)\citenamefont{Huang, Meng, Wang,
  Peng, Zhang, Chen, Li, Zhou, and Zhang}}]{Huang2016}
\bibinfo{author}{\bibfnamefont{L.}~\bibnamefont{Huang}},
  \bibinfo{author}{\bibfnamefont{Z.}~\bibnamefont{Meng}},
  \bibinfo{author}{\bibfnamefont{P.}~\bibnamefont{Wang}},
  \bibinfo{author}{\bibfnamefont{P.}~\bibnamefont{Peng}},
  \bibinfo{author}{\bibfnamefont{S.-L.} \bibnamefont{Zhang}},
  \bibinfo{author}{\bibfnamefont{L.}~\bibnamefont{Chen}},
  \bibinfo{author}{\bibfnamefont{D.}~\bibnamefont{Li}},
  \bibinfo{author}{\bibfnamefont{Q.}~\bibnamefont{Zhou}}, \bibnamefont{and}
  \bibinfo{author}{\bibfnamefont{J.}~\bibnamefont{Zhang}},
  \bibinfo{journal}{Nature Physics} \textbf{\bibinfo{volume}{12}},
  \bibinfo{pages}{540} (\bibinfo{year}{2016}).

\bibitem[{\citenamefont{Juzeliunas et~al.}(2005)\citenamefont{Juzeliunas,
  {\"{O}}hberg, Ruseckas, and Klein}}]{Juzeliunas2005}
\bibinfo{author}{\bibfnamefont{G.}~\bibnamefont{Juzeliunas}},
  \bibinfo{author}{\bibfnamefont{P.}~\bibnamefont{{\"{O}}hberg}},
  \bibinfo{author}{\bibfnamefont{J.}~\bibnamefont{Ruseckas}}, \bibnamefont{and}
  \bibinfo{author}{\bibfnamefont{A.}~\bibnamefont{Klein}},
  \bibinfo{journal}{Physical Review A} \textbf{\bibinfo{volume}{71}},
  \bibinfo{pages}{053614} (\bibinfo{year}{2005}).

\bibitem[{\citenamefont{Chen et~al.}(2018)\citenamefont{Chen, Lin, Chen, Chiu,
  Wang, Chen, Huang, Yip, Kawaguchi, and Lin}}]{Chen2018}
\bibinfo{author}{\bibfnamefont{H.-R.} \bibnamefont{Chen}},
  \bibinfo{author}{\bibfnamefont{K.-Y.} \bibnamefont{Lin}},
  \bibinfo{author}{\bibfnamefont{P.-K.} \bibnamefont{Chen}},
  \bibinfo{author}{\bibfnamefont{N.-C.} \bibnamefont{Chiu}},
  \bibinfo{author}{\bibfnamefont{J.-B.} \bibnamefont{Wang}},
  \bibinfo{author}{\bibfnamefont{C.-A.} \bibnamefont{Chen}},
  \bibinfo{author}{\bibfnamefont{P.-P.} \bibnamefont{Huang}},
  \bibinfo{author}{\bibfnamefont{S.-K.} \bibnamefont{Yip}},
  \bibinfo{author}{\bibfnamefont{Y.}~\bibnamefont{Kawaguchi}},
  \bibnamefont{and} \bibinfo{author}{\bibfnamefont{Y.-J.} \bibnamefont{Lin}},
  \bibinfo{journal}{Physical Review Letters} \textbf{\bibinfo{volume}{121}},
  \bibinfo{pages}{113204} (\bibinfo{year}{2018}).

\bibitem[{\citenamefont{Qu et~al.}(2015)\citenamefont{Qu, Sun, and
  Zhang}}]{Qu2015}
\bibinfo{author}{\bibfnamefont{C.}~\bibnamefont{Qu}},
  \bibinfo{author}{\bibfnamefont{K.}~\bibnamefont{Sun}}, \bibnamefont{and}
  \bibinfo{author}{\bibfnamefont{C.}~\bibnamefont{Zhang}},
  \bibinfo{journal}{Physical Review A} \textbf{\bibinfo{volume}{91}},
  \bibinfo{pages}{053630} (\bibinfo{year}{2015}).

\bibitem[{\citenamefont{DeMarco and Pu}(2015)}]{DeMarco2015}
\bibinfo{author}{\bibfnamefont{M.}~\bibnamefont{DeMarco}} \bibnamefont{and}
  \bibinfo{author}{\bibfnamefont{H.}~\bibnamefont{Pu}},
  \bibinfo{journal}{Physical Review A} \textbf{\bibinfo{volume}{91}},
  \bibinfo{pages}{033630} (\bibinfo{year}{2015}).

\bibitem[{\citenamefont{Hu et~al.}(2015)\citenamefont{Hu, Miniatura, and
  Gr{\'{e}}maud}}]{Hu2015}
\bibinfo{author}{\bibfnamefont{Y.-X.} \bibnamefont{Hu}},
  \bibinfo{author}{\bibfnamefont{C.}~\bibnamefont{Miniatura}},
  \bibnamefont{and}
  \bibinfo{author}{\bibfnamefont{B.}~\bibnamefont{Gr{\'{e}}maud}},
  \bibinfo{journal}{Physical Review A} \textbf{\bibinfo{volume}{92}},
  \bibinfo{pages}{033615} (\bibinfo{year}{2015}).

\bibitem[{\citenamefont{Chen et~al.}(2016)\citenamefont{Chen, Pu, and
  Zhang}}]{Chen2016}
\bibinfo{author}{\bibfnamefont{L.}~\bibnamefont{Chen}},
  \bibinfo{author}{\bibfnamefont{H.}~\bibnamefont{Pu}}, \bibnamefont{and}
  \bibinfo{author}{\bibfnamefont{Y.}~\bibnamefont{Zhang}},
  \bibinfo{journal}{Physical Review A} \textbf{\bibinfo{volume}{93}},
  \bibinfo{pages}{013629} (\bibinfo{year}{2016}).

\bibitem[{\citenamefont{Ramanathan et~al.}(2011)\citenamefont{Ramanathan,
  Wright, Muniz, Zelan, Hill, Lobb, Helmerson, Phillips, and
  Campbell}}]{Ramanathan2011}
\bibinfo{author}{\bibfnamefont{A.}~\bibnamefont{Ramanathan}},
  \bibinfo{author}{\bibfnamefont{K.~C.} \bibnamefont{Wright}},
  \bibinfo{author}{\bibfnamefont{S.~R.} \bibnamefont{Muniz}},
  \bibinfo{author}{\bibfnamefont{M.}~\bibnamefont{Zelan}},
  \bibinfo{author}{\bibfnamefont{W.~T.} \bibnamefont{Hill}},
  \bibinfo{author}{\bibfnamefont{C.~J.} \bibnamefont{Lobb}},
  \bibinfo{author}{\bibfnamefont{K.}~\bibnamefont{Helmerson}},
  \bibinfo{author}{\bibfnamefont{W.~D.} \bibnamefont{Phillips}},
  \bibnamefont{and} \bibinfo{author}{\bibfnamefont{G.~K.}
  \bibnamefont{Campbell}}, \bibinfo{journal}{Physical Review Letters}
  \textbf{\bibinfo{volume}{106}}, \bibinfo{pages}{130401}
  (\bibinfo{year}{2011}).

\bibitem[{\citenamefont{Beattie et~al.}(2013)\citenamefont{Beattie, Moulder,
  Fletcher, and Hadzibabic}}]{Beattie2013}
\bibinfo{author}{\bibfnamefont{S.}~\bibnamefont{Beattie}},
  \bibinfo{author}{\bibfnamefont{S.}~\bibnamefont{Moulder}},
  \bibinfo{author}{\bibfnamefont{R.~J.} \bibnamefont{Fletcher}},
  \bibnamefont{and}
  \bibinfo{author}{\bibfnamefont{Z.}~\bibnamefont{Hadzibabic}},
  \bibinfo{journal}{Physical Review Letters} \textbf{\bibinfo{volume}{110}},
  \bibinfo{pages}{025301} (\bibinfo{year}{2013}).

\bibitem[{\citenamefont{Wright et~al.}(2013)\citenamefont{Wright, Blakestad,
  Lobb, Phillips, and Campbell}}]{Wright2013}
\bibinfo{author}{\bibfnamefont{K.~C.} \bibnamefont{Wright}},
  \bibinfo{author}{\bibfnamefont{R.~B.} \bibnamefont{Blakestad}},
  \bibinfo{author}{\bibfnamefont{C.~J.} \bibnamefont{Lobb}},
  \bibinfo{author}{\bibfnamefont{W.~D.} \bibnamefont{Phillips}},
  \bibnamefont{and} \bibinfo{author}{\bibfnamefont{G.~K.}
  \bibnamefont{Campbell}}, \bibinfo{journal}{Physical Review Letters}
  \textbf{\bibinfo{volume}{110}}, \bibinfo{pages}{025302}
  (\bibinfo{year}{2013}).

\bibitem[{\citenamefont{Cooper and Hadzibabic}(2010)}]{Cooper2010}
\bibinfo{author}{\bibfnamefont{N.~R.} \bibnamefont{Cooper}} \bibnamefont{and}
  \bibinfo{author}{\bibfnamefont{Z.}~\bibnamefont{Hadzibabic}},
  \bibinfo{journal}{Physical Review Letters} \textbf{\bibinfo{volume}{104}},
  \bibinfo{pages}{030401} (\bibinfo{year}{2010}).

\bibitem[{\citenamefont{Leggett}(1999)}]{Leggett1999}
\bibinfo{author}{\bibfnamefont{A.~J.} \bibnamefont{Leggett}},
  \bibinfo{journal}{Reviews of Modern Physics} \textbf{\bibinfo{volume}{71}},
  \bibinfo{pages}{S318} (\bibinfo{year}{1999}).

\bibitem[{\citenamefont{Leggett}(2001)}]{Leggett2001}
\bibinfo{author}{\bibfnamefont{A.~J.} \bibnamefont{Leggett}},
  \bibinfo{journal}{Reviews of Modern Physics} \textbf{\bibinfo{volume}{73}},
  \bibinfo{pages}{307} (\bibinfo{year}{2001}).

\bibitem[{\citenamefont{Hess and Fairbank}(1967)}]{Hess1967}
\bibinfo{author}{\bibfnamefont{G.~B.} \bibnamefont{Hess}} \bibnamefont{and}
  \bibinfo{author}{\bibfnamefont{W.~M.} \bibnamefont{Fairbank}},
  \bibinfo{journal}{Physical Review Letters} \textbf{\bibinfo{volume}{19}},
  \bibinfo{pages}{216} (\bibinfo{year}{1967}).

\bibitem[{\citenamefont{Ishiguro et~al.}(2004)\citenamefont{Ishiguro, Ishikawa,
  Yamashita, Sasaki, Fukuda, Kubota, Ishimoto, Packard, Takagi, Ohmi
  et~al.}}]{Ishiguro2004}
\bibinfo{author}{\bibfnamefont{R.}~\bibnamefont{Ishiguro}},
  \bibinfo{author}{\bibfnamefont{O.}~\bibnamefont{Ishikawa}},
  \bibinfo{author}{\bibfnamefont{M.}~\bibnamefont{Yamashita}},
  \bibinfo{author}{\bibfnamefont{Y.}~\bibnamefont{Sasaki}},
  \bibinfo{author}{\bibfnamefont{K.}~\bibnamefont{Fukuda}},
  \bibinfo{author}{\bibfnamefont{M.}~\bibnamefont{Kubota}},
  \bibinfo{author}{\bibfnamefont{H.}~\bibnamefont{Ishimoto}},
  \bibinfo{author}{\bibfnamefont{R.~E.} \bibnamefont{Packard}},
  \bibinfo{author}{\bibfnamefont{T.}~\bibnamefont{Takagi}},
  \bibinfo{author}{\bibfnamefont{T.}~\bibnamefont{Ohmi}}, \bibnamefont{et~al.},
  \bibinfo{journal}{Physical Review Letters} \textbf{\bibinfo{volume}{93}},
  \bibinfo{pages}{125301} (\bibinfo{year}{2004}).

\bibitem[{\citenamefont{Moulder}(2013)}]{Moulder2013}
\bibinfo{author}{\bibfnamefont{S.}~\bibnamefont{Moulder}}, Ph.D. thesis,
  \bibinfo{school}{University of Cambridge} (\bibinfo{year}{2013}).

\bibitem[{\citenamefont{Kawaguchi and Ueda}(2012)}]{Kawaguchi2012}
\bibinfo{author}{\bibfnamefont{Y.}~\bibnamefont{Kawaguchi}} \bibnamefont{and}
  \bibinfo{author}{\bibfnamefont{M.}~\bibnamefont{Ueda}},
  \bibinfo{journal}{Physics Reports} \textbf{\bibinfo{volume}{520}},
  \bibinfo{pages}{253} (\bibinfo{year}{2012}).

\bibitem[{\citenamefont{Ueda}(2014)}]{Ueda2014}
\bibinfo{author}{\bibfnamefont{M.}~\bibnamefont{Ueda}},
  \bibinfo{journal}{Reports on Progress in Physics}
  \textbf{\bibinfo{volume}{77}} (\bibinfo{year}{2014}).

\bibitem[{\citenamefont{Isoshima et~al.}(2000)\citenamefont{Isoshima, Nakahara,
  Ohmi, and Machida}}]{Isoshima2000}
\bibinfo{author}{\bibfnamefont{T.}~\bibnamefont{Isoshima}},
  \bibinfo{author}{\bibfnamefont{M.}~\bibnamefont{Nakahara}},
  \bibinfo{author}{\bibfnamefont{T.}~\bibnamefont{Ohmi}}, \bibnamefont{and}
  \bibinfo{author}{\bibfnamefont{K.}~\bibnamefont{Machida}},
  \bibinfo{journal}{Physical Review A} \textbf{\bibinfo{volume}{61}},
  \bibinfo{pages}{063610} (\bibinfo{year}{2000}).

\bibitem[{\citenamefont{Leanhardt et~al.}(2003)\citenamefont{Leanhardt, Shin,
  Kielpinski, Pritchard, and Ketterle}}]{Leanhardt2003}
\bibinfo{author}{\bibfnamefont{A.~E.} \bibnamefont{Leanhardt}},
  \bibinfo{author}{\bibfnamefont{Y.}~\bibnamefont{Shin}},
  \bibinfo{author}{\bibfnamefont{D.}~\bibnamefont{Kielpinski}},
  \bibinfo{author}{\bibfnamefont{D.~E.} \bibnamefont{Pritchard}},
  \bibnamefont{and} \bibinfo{author}{\bibfnamefont{W.}~\bibnamefont{Ketterle}},
  \bibinfo{journal}{Physical Review Letters} \textbf{\bibinfo{volume}{90}},
  \bibinfo{pages}{140403} (\bibinfo{year}{2003}).

\bibitem[{\citenamefont{Ray et~al.}(2014)\citenamefont{Ray, Ruokokoski, Kandel,
  M{\"{o}}tt{\"{o}}nen, and Hall}}]{Ray2014}
\bibinfo{author}{\bibfnamefont{M.~W.} \bibnamefont{Ray}},
  \bibinfo{author}{\bibfnamefont{E.}~\bibnamefont{Ruokokoski}},
  \bibinfo{author}{\bibfnamefont{S.}~\bibnamefont{Kandel}},
  \bibinfo{author}{\bibfnamefont{M.}~\bibnamefont{M{\"{o}}tt{\"{o}}nen}},
  \bibnamefont{and} \bibinfo{author}{\bibfnamefont{D.~S.} \bibnamefont{Hall}},
  \bibinfo{journal}{Nature} \textbf{\bibinfo{volume}{505}},
  \bibinfo{pages}{657} (\bibinfo{year}{2014}).

\bibitem[{\citenamefont{Ray et~al.}(2015)\citenamefont{Ray, Ruokokoski, Tiurev,
  M{\"{o}}tt{\"{o}}nen, and Hall}}]{Ray2015}
\bibinfo{author}{\bibfnamefont{M.~W.} \bibnamefont{Ray}},
  \bibinfo{author}{\bibfnamefont{E.}~\bibnamefont{Ruokokoski}},
  \bibinfo{author}{\bibfnamefont{K.}~\bibnamefont{Tiurev}},
  \bibinfo{author}{\bibfnamefont{M.}~\bibnamefont{M{\"{o}}tt{\"{o}}nen}},
  \bibnamefont{and} \bibinfo{author}{\bibfnamefont{D.~S.} \bibnamefont{Hall}},
  \bibinfo{journal}{Science} \textbf{\bibinfo{volume}{348}},
  \bibinfo{pages}{544} (\bibinfo{year}{2015}).

\bibitem[{\citenamefont{Hall et~al.}(2016)\citenamefont{Hall, Ray, Tiurev,
  Ruokokoski, Gheorghe, and M{\"{o}}tt{\"{o}}nen}}]{Hall2016}
\bibinfo{author}{\bibfnamefont{D.~S.} \bibnamefont{Hall}},
  \bibinfo{author}{\bibfnamefont{M.~W.} \bibnamefont{Ray}},
  \bibinfo{author}{\bibfnamefont{K.}~\bibnamefont{Tiurev}},
  \bibinfo{author}{\bibfnamefont{E.}~\bibnamefont{Ruokokoski}},
  \bibinfo{author}{\bibfnamefont{A.~H.} \bibnamefont{Gheorghe}},
  \bibnamefont{and}
  \bibinfo{author}{\bibfnamefont{M.}~\bibnamefont{M{\"{o}}tt{\"{o}}nen}},
  \bibinfo{journal}{Nature Physics} \textbf{\bibinfo{volume}{12}},
  \bibinfo{pages}{478} (\bibinfo{year}{2016}).

\bibitem[{\citenamefont{Ollikainen et~al.}(2017)\citenamefont{Ollikainen,
  Tiurev, Blinova, Lee, Hall, and M{\"{o}}tt{\"{o}}nen}}]{Ollikainen2017}
\bibinfo{author}{\bibfnamefont{T.}~\bibnamefont{Ollikainen}},
  \bibinfo{author}{\bibfnamefont{K.}~\bibnamefont{Tiurev}},
  \bibinfo{author}{\bibfnamefont{A.}~\bibnamefont{Blinova}},
  \bibinfo{author}{\bibfnamefont{W.}~\bibnamefont{Lee}},
  \bibinfo{author}{\bibfnamefont{D.}~\bibnamefont{Hall}}, \bibnamefont{and}
  \bibinfo{author}{\bibfnamefont{M.}~\bibnamefont{M{\"{o}}tt{\"{o}}nen}},
  \bibinfo{journal}{Physical Review X} \textbf{\bibinfo{volume}{7}},
  \bibinfo{pages}{021023} (\bibinfo{year}{2017}).

\bibitem[{\citenamefont{Choi et~al.}(2012{\natexlab{a}})\citenamefont{Choi,
  Kwon, and Shin}}]{Choi2012a}
\bibinfo{author}{\bibfnamefont{J.-Y.} \bibnamefont{Choi}},
  \bibinfo{author}{\bibfnamefont{W.~J.} \bibnamefont{Kwon}}, \bibnamefont{and}
  \bibinfo{author}{\bibfnamefont{Y.-I.} \bibnamefont{Shin}},
  \bibinfo{journal}{Physical Review Letters} \textbf{\bibinfo{volume}{108}},
  \bibinfo{pages}{035301} (\bibinfo{year}{2012}{\natexlab{a}}).

\bibitem[{\citenamefont{Choi et~al.}(2012{\natexlab{b}})\citenamefont{Choi,
  Kwon, Lee, Jeong, An, and Shin}}]{Choi2012}
\bibinfo{author}{\bibfnamefont{J.-Y.} \bibnamefont{Choi}},
  \bibinfo{author}{\bibfnamefont{W.~J.} \bibnamefont{Kwon}},
  \bibinfo{author}{\bibfnamefont{M.}~\bibnamefont{Lee}},
  \bibinfo{author}{\bibfnamefont{H.}~\bibnamefont{Jeong}},
  \bibinfo{author}{\bibfnamefont{K.}~\bibnamefont{An}}, \bibnamefont{and}
  \bibinfo{author}{\bibfnamefont{Y.-i.} \bibnamefont{Shin}},
  \bibinfo{journal}{New Journal of Physics} \textbf{\bibinfo{volume}{14}},
  \bibinfo{pages}{053013} (\bibinfo{year}{2012}{\natexlab{b}}).

\bibitem[{\citenamefont{Lee et~al.}(2018)\citenamefont{Lee, Gheorghe, Tiurev,
  Ollikainen, M{\"{o}}tt{\"{o}}nen, and Hall}}]{Lee2018}
\bibinfo{author}{\bibfnamefont{W.}~\bibnamefont{Lee}},
  \bibinfo{author}{\bibfnamefont{A.~H.} \bibnamefont{Gheorghe}},
  \bibinfo{author}{\bibfnamefont{K.}~\bibnamefont{Tiurev}},
  \bibinfo{author}{\bibfnamefont{T.}~\bibnamefont{Ollikainen}},
  \bibinfo{author}{\bibfnamefont{M.}~\bibnamefont{M{\"{o}}tt{\"{o}}nen}},
  \bibnamefont{and} \bibinfo{author}{\bibfnamefont{D.~S.} \bibnamefont{Hall}},
  \bibinfo{journal}{Science Advances} \textbf{\bibinfo{volume}{4}},
  \bibinfo{pages}{eaao3820} (\bibinfo{year}{2018}).

\bibitem[{\citenamefont{Choi et~al.}(2013)\citenamefont{Choi, Kang, Seo, Kwon,
  and Shin}}]{Choi2013}
\bibinfo{author}{\bibfnamefont{J.-Y.} \bibnamefont{Choi}},
  \bibinfo{author}{\bibfnamefont{S.}~\bibnamefont{Kang}},
  \bibinfo{author}{\bibfnamefont{S.~W.} \bibnamefont{Seo}},
  \bibinfo{author}{\bibfnamefont{W.~J.} \bibnamefont{Kwon}}, \bibnamefont{and}
  \bibinfo{author}{\bibfnamefont{Y.-i.} \bibnamefont{Shin}},
  \bibinfo{journal}{Physical Review Letters} \textbf{\bibinfo{volume}{111}},
  \bibinfo{pages}{245301} (\bibinfo{year}{2013}).

\bibitem[{\citenamefont{Leslie et~al.}(2009)\citenamefont{Leslie, Hansen,
  Wright, Deutsch, and Bigelow}}]{Leslie2009}
\bibinfo{author}{\bibfnamefont{L.~S.} \bibnamefont{Leslie}},
  \bibinfo{author}{\bibfnamefont{A.}~\bibnamefont{Hansen}},
  \bibinfo{author}{\bibfnamefont{K.~C.} \bibnamefont{Wright}},
  \bibinfo{author}{\bibfnamefont{B.~M.} \bibnamefont{Deutsch}},
  \bibnamefont{and} \bibinfo{author}{\bibfnamefont{N.~P.}
  \bibnamefont{Bigelow}}, \bibinfo{journal}{Physical Review Letters}
  \textbf{\bibinfo{volume}{103}}, \bibinfo{pages}{250401}
  (\bibinfo{year}{2009}).

\bibitem[{\citenamefont{Hansen}(2016)}]{Hansen2016}
\bibinfo{author}{\bibfnamefont{A.}~\bibnamefont{Hansen}}, Ph.D. thesis,
  \bibinfo{school}{University of Rochester} (\bibinfo{year}{2016}).

\bibitem[{\citenamefont{Isoshima et~al.}(2001)\citenamefont{Isoshima, Machida,
  and Ohmi}}]{Isoshima2001}
\bibinfo{author}{\bibfnamefont{T.}~\bibnamefont{Isoshima}},
  \bibinfo{author}{\bibfnamefont{K.}~\bibnamefont{Machida}}, \bibnamefont{and}
  \bibinfo{author}{\bibfnamefont{T.}~\bibnamefont{Ohmi}},
  \bibinfo{journal}{Journal of the Physical Society of Japan}
  \textbf{\bibinfo{volume}{70}}, \bibinfo{pages}{1604} (\bibinfo{year}{2001}).

\bibitem[{\citenamefont{Sadler et~al.}(2006)\citenamefont{Sadler, Higbie,
  Leslie, Vengalattore, and Stamper-Kurn}}]{Sadler2006}
\bibinfo{author}{\bibfnamefont{L.~E.} \bibnamefont{Sadler}},
  \bibinfo{author}{\bibfnamefont{J.~M.} \bibnamefont{Higbie}},
  \bibinfo{author}{\bibfnamefont{S.~R.} \bibnamefont{Leslie}},
  \bibinfo{author}{\bibfnamefont{M.}~\bibnamefont{Vengalattore}},
  \bibnamefont{and} \bibinfo{author}{\bibfnamefont{D.~M.}
  \bibnamefont{Stamper-Kurn}}, \bibinfo{journal}{Nature}
  \textbf{\bibinfo{volume}{443}}, \bibinfo{pages}{312} (\bibinfo{year}{2006}).

\bibitem[{\citenamefont{Ho}(1998)}]{Ho1998}
\bibinfo{author}{\bibfnamefont{T.-L.} \bibnamefont{Ho}},
  \bibinfo{journal}{Physical Review Letters} \textbf{\bibinfo{volume}{81}},
  \bibinfo{pages}{742} (\bibinfo{year}{1998}).

\bibitem[{\citenamefont{Ho and Shenoy}(1996)}]{Ho1996}
\bibinfo{author}{\bibfnamefont{T.-L.} \bibnamefont{Ho}} \bibnamefont{and}
  \bibinfo{author}{\bibfnamefont{V.~B.} \bibnamefont{Shenoy}},
  \bibinfo{journal}{Physical Review Letters} \textbf{\bibinfo{volume}{77}},
  \bibinfo{pages}{2595} (\bibinfo{year}{1996}).

\bibitem[{\citenamefont{Castin and Dum}(1996)}]{Castin1996}
\bibinfo{author}{\bibfnamefont{Y.}~\bibnamefont{Castin}} \bibnamefont{and}
  \bibinfo{author}{\bibfnamefont{R.}~\bibnamefont{Dum}},
  \bibinfo{journal}{Physical Review Letters} \textbf{\bibinfo{volume}{77}},
  \bibinfo{pages}{5315} (\bibinfo{year}{1996}).

\bibitem[{\citenamefont{Haljan et~al.}(2001)\citenamefont{Haljan, Coddington,
  Engels, and Cornell}}]{Haljan2001}
\bibinfo{author}{\bibfnamefont{P.~C.} \bibnamefont{Haljan}},
  \bibinfo{author}{\bibfnamefont{I.}~\bibnamefont{Coddington}},
  \bibinfo{author}{\bibfnamefont{P.}~\bibnamefont{Engels}}, \bibnamefont{and}
  \bibinfo{author}{\bibfnamefont{E.~A.} \bibnamefont{Cornell}},
  \bibinfo{journal}{Physical Review Letters} \textbf{\bibinfo{volume}{87}},
  \bibinfo{pages}{210403} (\bibinfo{year}{2001}).

\bibitem[{\citenamefont{Madison et~al.}(2001)\citenamefont{Madison, Chevy,
  Bretin, and Dalibard}}]{madison01}
\bibinfo{author}{\bibfnamefont{K.~W.} \bibnamefont{Madison}},
  \bibinfo{author}{\bibfnamefont{F.}~\bibnamefont{Chevy}},
  \bibinfo{author}{\bibfnamefont{V.}~\bibnamefont{Bretin}}, \bibnamefont{and}
  \bibinfo{author}{\bibfnamefont{J.}~\bibnamefont{Dalibard}},
  \bibinfo{journal}{Phys. Rev. Lett.} \textbf{\bibinfo{volume}{86}},
  \bibinfo{pages}{4443} (\bibinfo{year}{2001}).

\bibitem[{\citenamefont{Abo-Shaeer et~al.}(2001)\citenamefont{Abo-Shaeer,
  Raman, Vogels, and Ketterle}}]{Aboshaeer01}
\bibinfo{author}{\bibfnamefont{J.~R.} \bibnamefont{Abo-Shaeer}},
  \bibinfo{author}{\bibfnamefont{C.}~\bibnamefont{Raman}},
  \bibinfo{author}{\bibfnamefont{J.~M.} \bibnamefont{Vogels}},
  \bibnamefont{and} \bibinfo{author}{\bibfnamefont{W.}~\bibnamefont{Ketterle}},
  \bibinfo{journal}{Science} \textbf{\bibinfo{volume}{292}},
  \bibinfo{pages}{476} (\bibinfo{year}{2001}).

\bibitem[{\citenamefont{Hodby et~al.}(2001)\citenamefont{Hodby, Hechenblaikner,
  Hopkins, Marag{\`{o}}, and Foot}}]{Hodby2001}
\bibinfo{author}{\bibfnamefont{E.}~\bibnamefont{Hodby}},
  \bibinfo{author}{\bibfnamefont{G.}~\bibnamefont{Hechenblaikner}},
  \bibinfo{author}{\bibfnamefont{S.~A.} \bibnamefont{Hopkins}},
  \bibinfo{author}{\bibfnamefont{O.~M.} \bibnamefont{Marag{\`{o}}}},
  \bibnamefont{and} \bibinfo{author}{\bibfnamefont{C.~J.} \bibnamefont{Foot}},
  \bibinfo{journal}{Physical Review Letters} \textbf{\bibinfo{volume}{88}},
  \bibinfo{pages}{010405} (\bibinfo{year}{2001}).

\bibitem[{\citenamefont{Lundh et~al.}(1997)\citenamefont{Lundh, Pethick, and
  Smith}}]{Lundh1997}
\bibinfo{author}{\bibfnamefont{E.}~\bibnamefont{Lundh}},
  \bibinfo{author}{\bibfnamefont{C.~J.} \bibnamefont{Pethick}},
  \bibnamefont{and} \bibinfo{author}{\bibfnamefont{H.}~\bibnamefont{Smith}},
  \bibinfo{journal}{Physical Review A} \textbf{\bibinfo{volume}{55}},
  \bibinfo{pages}{2126} (\bibinfo{year}{1997}).

\bibitem[{\citenamefont{Juzeliu¯nas and {\"{O}}hberg}(2005)}]{Juzeliunas2005a}
\bibinfo{author}{\bibfnamefont{G.}~\bibnamefont{Juzeliunas}}
  \bibnamefont{and}
  \bibinfo{author}{\bibfnamefont{P.}~\bibnamefont{{\"{O}}hberg}},
  \bibinfo{journal}{Optics and Spectroscopy} \textbf{\bibinfo{volume}{99}},
  \bibinfo{pages}{357} (\bibinfo{year}{2005}).

\bibitem[{\citenamefont{Zhang et~al.}(2018)\citenamefont{Zhang, Gao, Zou, Kong,
  Li, Shen, Chen, Peng, Zhan, Pu et~al.}}]{Zhang2018}
\bibinfo{author}{\bibfnamefont{D.}~\bibnamefont{Zhang}},
  \bibinfo{author}{\bibfnamefont{T.}~\bibnamefont{Gao}},
  \bibinfo{author}{\bibfnamefont{P.}~\bibnamefont{Zou}},
  \bibinfo{author}{\bibfnamefont{L.}~\bibnamefont{Kong}},
  \bibinfo{author}{\bibfnamefont{R.}~\bibnamefont{Li}},
  \bibinfo{author}{\bibfnamefont{X.}~\bibnamefont{Shen}},
  \bibinfo{author}{\bibfnamefont{X.-L.} \bibnamefont{Chen}},
  \bibinfo{author}{\bibfnamefont{S.-G.} \bibnamefont{Peng}},
  \bibinfo{author}{\bibfnamefont{M.}~\bibnamefont{Zhan}},
  \bibinfo{author}{\bibfnamefont{H.}~\bibnamefont{Pu}}, \bibnamefont{et~al.}
  (\bibinfo{year}{2018}), \eprint{arXiv:1806.06263}.

\end{thebibliography}

\begin{thebibliography}{2}
\expandafter\ifx\csname natexlab\endcsname\relax\def\natexlab#1{#1}\fi
\expandafter\ifx\csname bibnamefont\endcsname\relax
  \def\bibnamefont#1{#1}\fi
\expandafter\ifx\csname bibfnamefont\endcsname\relax
  \def\bibfnamefont#1{#1}\fi
\expandafter\ifx\csname citenamefont\endcsname\relax
  \def\citenamefont#1{#1}\fi
\expandafter\ifx\csname url\endcsname\relax
  \def\url#1{\texttt{#1}}\fi
\expandafter\ifx\csname urlprefix\endcsname\relax\def\urlprefix{URL }\fi
\providecommand{\bibinfo}[2]{#2}
\providecommand{\eprint}[2][]{\url{#2}}

\bibitem[{\citenamefont{Chen et~al.}(2018)\citenamefont{Chen, Lin, Chen, Chiu,
  Wang, Chen, Huang, Yip, Kawaguchi, and Lin}}]{SChen2018}
\bibinfo{author}{\bibfnamefont{H.~R.} \bibnamefont{Chen}},
  \bibinfo{author}{\bibfnamefont{K.~Y.} \bibnamefont{Lin}},
  \bibinfo{author}{\bibfnamefont{P.~K.} \bibnamefont{Chen}},
  \bibinfo{author}{\bibfnamefont{N.~C.} \bibnamefont{Chiu}},
  \bibinfo{author}{\bibfnamefont{J.~B.} \bibnamefont{Wang}},
  \bibinfo{author}{\bibfnamefont{C.~A.} \bibnamefont{Chen}},
  \bibinfo{author}{\bibfnamefont{P.~P.} \bibnamefont{Huang}},
  \bibinfo{author}{\bibfnamefont{S.~K.} \bibnamefont{Yip}},
  \bibinfo{author}{\bibfnamefont{Y.}~\bibnamefont{Kawaguchi}},
  \bibnamefont{and} \bibinfo{author}{\bibfnamefont{Y.~J.} \bibnamefont{Lin}}
  (\bibinfo{year}{2018}), \eprint{arXiv:1803.07860}.

\bibitem[{\citenamefont{Reinaudi et~al.}(2007)\citenamefont{Reinaudi, Lahaye,
  Wang, and Gu{\'{e}}ry-Odelin}}]{SReinaudi2007}
\bibinfo{author}{\bibfnamefont{G.}~\bibnamefont{Reinaudi}},
  \bibinfo{author}{\bibfnamefont{T.}~\bibnamefont{Lahaye}},
  \bibinfo{author}{\bibfnamefont{Z.}~\bibnamefont{Wang}}, \bibnamefont{and}
  \bibinfo{author}{\bibfnamefont{D.}~\bibnamefont{Gu{\'{e}}ry-Odelin}},
  \bibinfo{journal}{Optics Letters} \textbf{\bibinfo{volume}{32}},
  \bibinfo{pages}{3143} (\bibinfo{year}{2007}).

\bibitem[{\citenamefont{Ho}(1998)}]{SHo1998}
\bibinfo{author}{\bibfnamefont{T.-L.} \bibnamefont{Ho}},
  \bibinfo{journal}{Physical Review Letters} \textbf{\bibinfo{volume}{81}},
  \bibinfo{pages}{742} (\bibinfo{year}{1998}).

\bibitem[{\citenamefont{Lundh et~al.}(1997)\citenamefont{Lundh, Pethick, and
  Smith}}]{SLundh1997}
\bibinfo{author}{\bibfnamefont{E.}~\bibnamefont{Lundh}},
  \bibinfo{author}{\bibfnamefont{C.~J.} \bibnamefont{Pethick}},
  \bibnamefont{and} \bibinfo{author}{\bibfnamefont{H.}~\bibnamefont{Smith}},
  \bibinfo{journal}{Physical Review A} \textbf{\bibinfo{volume}{55}},
  \bibinfo{pages}{2126} (\bibinfo{year}{1997}).

\end{thebibliography}
\end{document}